\documentclass[12pt,reqno]{article}
\usepackage{latexsym,amsmath,amsthm,amsfonts,amssymb,amscd}
\usepackage[dvips]{graphicx,psfrag}

\def\P {{\mathbf P}}

\def\QQ {{\mathbb Q}}
\def\RR {{\mathbb R}}

\def\cE{{\cal E}}

\newtheorem{lemm}{Lemma}[section]

\newtheorem{defi}[lemm]{Definition}

\title{\bf The Fundamental Concepts  of Classical
Equilibrium Statistical Mechanics }

\author{S\'ergio B. Volchan\thanks{Pontif\'{\i}cia Universidade Cat\'olica do 
Rio de Janeiro, Departamento de Matem\'atica, Rua Marqu\^es de S\~ao Vicente 
225,G\'avea, 22453-900 Rio de Janeiro, Brasil
volchan@mat.puc-rio.br}}

\makeatletter
\renewcommand{\@biblabel}[1]{#1.}
\renewcommand{\@cite}[2]{$^{({#1\if@tempswa, #2\fi})}$}
\makeatother

\begin{document}
\vskip .5cm

\date{}
\maketitle

\vskip .5cm
\begin{flushright}
\end{flushright}
\vskip .8cm

\begin{abstract}
A critical examination  of some basic conceptual issues 
in classical statistical mechanics is attempted, 
with a view to understanding  the origins, structure and status of that
discipline. Due attention is  given to the interplay between
physical and mathematical aspects, particularly regarding the role of  
probability theory. The focus is on the equilibrium case, which is currently 
better understood, serving also  as a prelude for a further discussion  
non-equilibrium statistical mechanics.
\end{abstract}




\baselineskip 12pt
 
\noindent\section{Introduction} 

It is a striking feature of the world that it has a multilevel structure.
From subatomic particles to galaxies, there is a great variety of levels of
reality,  each with its own objects, properties and laws. The effort in dealing
with such richness is reflected in the division of labor of the scientific 
enterprise,  each  discipline trying to map and understand some part of the 
complex whole. 

Though the aforementioned  levels are autonomous to a great extent,  
they are not totally  independent. Therefore, once a reasonable understanding 
of phenomena at some of these levels is accomplished, there naturally arises 
the task of an {\em inter}level investigation. It should address questions 
such as: how are levels organized with respect to each other, 
is there a natural hierarchy or structure of levels, how do new properties 
emerge from ``lower'' to ``higher'' levels, how can one explain higher levels 
in terms of the lower ones, etc? One could fairly say that the elucidation of 
the connections  among levels of reality is a major test of the coherence of 
the  scientific worldview and, besides functioning  as a fine tuning for our 
theories, such a study not infrequently lead to new discoveries and further 
inquiries.

Now, one of the earliest and broadest level distinctions, of particular
importance to physics, is that between the so-called {\em macroscopic} and 
{\em microscopic} levels. It stems from the  notion that, 
underlying the world of the visible and apparently homogeneous substances, 
there is a more ``basic'' reality consisting of a very large number of tiny 
invisible (and indivisible) discrete components.~\footnote{More generally, it 
corresponds to the notion that a necessary aspect of any {\em system}
is that it has components.~\cite{Bun2}}  

In principle, the microreality would be considered more basic in the sense 
that the directly observable phenomena would result from (or be explained by), 
the complicated motions and mutual arrangements of those components. This is 
essentially the ``atomic hypothesis'' (or atomism)  which, together with  
mechanics and probability theory, are the  main ingredients 
from which  statistical mechanics emerged  in  the last half of the 
XIX$^{\mbox{\scriptsize th}}$ century and the first decades of the  
XX$^{\mbox{\scriptsize th}}$, out of the efforts to provide a 
mechanical-atomistic foundation of thermodynamics.

Statistical mechanics can then be conceived of as a discipline (or, 
maybe, a set of techniques and prescriptions) whose aim is to 
serve as a {\em bridge} between the micro and macro levels. In its role
as a {\em level-connecting} discipline, it acquired a peculiar flavor. 
So, in spite of having appeared  in the somewhat narrow context of
the study of gases, it is supposed to be very general to the point of being 
a sort of ``super-theory''; for example it was  instrumental 
in the advent of the quantum revolution, more specifically in
Planck's 1900 solution of the black-body radiation conundrum. Its
ideas and techniques are frequently used (and sometimes abused) in
such disparate  areas as quantum field theory, turbulence, dynamical systems,
image processing, neural networks, computational complexity theory, 
biology and finance. This is certainly linked to the pivotal 
role  of probability theory, with its very  general notions and theorems,
in the framework of statistical mechanics.

Also,  the mathematically rigorous analysis of specific 
statistical mechanical systems proposed in the physics literature turned
out to be very difficult, even for some highly idealized models, like 
lattice gases. So statistical mechanics became also the battle ground
par excellence for mathematical-physics, inspiring the creation of new 
concepts and techniques to deal with its problems. We think that 
statistical mechanics clearly illustrates the inestimable role of
mathematical-physics  in bringing precision and organization to a 
notoriously difficult subject. It is also interesting to witness once more
how such rich, sophisticated and highly abstract mathematical machinery is 
needed even to formulate (not to mention solve)  statistical mechanical 
problems on a rigorous basis. In any case, statistical mechanics has proven 
to be an indispensable and extremely rich tool of research in many-body 
physics, presenting  many hard questions of physical, mathematical, 
conceptual, methodological and philosophical importance.~\cite{Ruelle2}

In this paper we intend to examine  only a sample of issues in this
already vast field, hoping to contribute to a better 
understanding of its role, structure and methods. We will focus mainly on 
fundamental concepts which seem to be at its core. Due attention is  
payed to the interplay between the physical-conceptual problems and the 
corresponding mathematical ideas, methods and theories used to formulate 
them in a rigorous fashion. 

We will be mainly concerned  with {\em classical equilibrium statistical 
mechanics}, leaving a discussion of the much more complicated (and more
interesting) case of non-equilibrium statistical mechanics 
(whatever that might be) to another occasion. Although the two 
branches are historically and inextricably  linked, the non-equilibrium case 
is, at the present stage of research, much less 
understood.  Accordingly, a common research strategy has been to adapt some 
concepts from the former in trying to come to terms with the 
latter.~\footnote{A case in point is  the important (and delicate) notion of 
local equilibrium in non-equilibrium statistical mechanics.}  In this sense, 
one can also say that an acquaintance with the equilibrium situation might
be a useful prerequisite to an understanding of non-equilibrium issues.

The paper is structured as follows. We first recall the main influences
in the emergence of statistical mechanics and which strongly shaped its 
subsequent development.  We then discuss with some detail
the basic notions of the ``ensemble'' theory. Finally, we
touch on the central and subtle topic of phase transitions, after which
we make some concluding remarks.

\noindent\section{Preliminaries}

Without delving into the fascinating and rather convoluted history of
the emergence of statistical mechanics,~\footnote{A history which is 
yet to be written. See however references 3, 12 and 16.} it is useful to
summarize the main influences in its inception. This will provide a
broader  context that helps one grasp the sources of its main problems, 
aims and methods.

\subsection{Thermodynamics}

The first (and historically crucial) ingredient is of course, 
thermodynamics. In fact, the very idea  of providing an atomistic-mechanical 
basis for it, can be taken as the point of  departure of the statistical 
mechanical ``program'' (for example, in the guise of kinetic gas theory). 

Thermodynamics, together with  classical  mechanics and electrodynamics, was  
one of the pillars of late nineteenth-century physics. It is an amazingly 
general phenomenological theory, concerning properties and processes of 
macroscopic systems  (typically continuum media such as gases and  fluids, 
but including reacting chemicals, magnetic systems, etc) regarding 
exchanges of  heat, energy and matter. As such, it is an  indispensable tool 
in many technological areas, particularly to engineering. 

Notwithstanding the traditional textbook view of thermodynamics 
as a completed (and even stagnant) discipline, it is actually a
very live research field, full of open  problems and some ongoing 
controversies.~\footnote{It is almost a scandal that one could 
complete  a graduate program in theoretical physics without 
realizing the existence of such controversies and/or its modern 
developments.} In particular, one observes a sharp distinction of methodology 
and conceptual viewpoints between  the rational-mechanics
community~\footnote{We refer to the school led by W. Noll, the late 
C. Truesdell,  J. Serrin and many others.~\cite{True}} and  the mainstream 
physics community.~\footnote{This curious (and unfortunate) lack of exchange 
between these research communities (and which deserved to be mended) would be 
an interesting  case study in the sociology of science.} 

It was the unsatisfactory state of standard presentations of thermodynamics
and the concomitant conceptual confusion, that has motivated the  many 
attempts at a clarification of its foundations.~\footnote{There is some
similarity between the situation of the foundations of thermodynamics 
(particularly regarding its  conceptual confusion),  with that  
of quantum mechanics. So, thermodynamics had a rather influential but 
unsuccessful axiomatization in Caratheod\'ory's (1909) work,~\cite{True2}
and the same can be said of von Neumann's  ill-fated  axiomatization of 
quantum mechanics (1932).~\cite{Wick} And, as it happened with the effort of 
clarification of thermodynamics, there has been recently an 
effort to reassess the foundations of quantum mechanics, for example, 
through a renewed  version of the much  neglected  Bohmian 
approach.~\cite{Golds}} Ideally, as suggested by David Hilbert, this 
conceptual elucidation should proceed through a careful
axiomatization of the theory.~\footnote{The sixth problem in his 
famous list of 23 problems, proposed in 1900 at the Second International
Congress of Mathematics in Paris, concerns the axiomatization of physical 
theories.~\cite{Corry}} At present, there is a variety of formulations,
with different degrees of rigor and generality, but still no
universal agreement.  However, this does not mean the effort is 
worthless. Quite on the contrary, it signals that thermodynamics is a 
difficult and subtle discipline in need of conceptual 
clarification.~\footnote{In particular, if one intends to deduce thermodynamics
from  a more basic microscopic theory, it would be desirable to have
a clear understanding and formalization of it.} A  
detailed critical review of the conceptual problems of thermodynamics is
beyond the scope of this paper and in the following  we limit ourselves to
some general comments (see also ref. 51 )

 The usual presentations of thermodynamics discuss the three 
fundamental laws,  starting from some basic concepts,~\footnote{Which, in an
axiomatic formulation, should figure among the primitive 
notions, that is, basic undefined concepts, a point is which rarely made 
explicit or even clearly discussed.} say,  of {\em  system}, {\em state} and 
{\em equilibrium}. A {\em thermodynamic  system} is characterized by its  
physico-chemical properties, like total mass and chemical composition, and
also by a (real or hypothetical) boundary  separating it from the 
environment with which it interacts. A system is {\em closed} when 
there is no exchange of matter, otherwise it is open. Usually the theory is
formulated for closed ones. Also, a (closed) system is 
{\em isolated} when it does not interact with the
exterior, i.e., there is no exchange of heat nor work is performed (it
can be conceived of as a system enclosed by rigid adiabatic walls).
 
The {\em thermodynamic state} of the system is usually specified by a 
relatively small number of  internal and external {\em parameters} 
(or {\em state variables}) (e.g., temperature, pressure,  volume, internal 
energy and density for gases  and fluids; magnetic field and  magnetization 
for magnetic systems) that completely characterize the system in 
{\em equilibrium}. The {\em equilibrium states} of each system are completely 
determined~\footnote{Except, possibly, in the presence of phase transitions, 
see section 4.} by a set of independent parameters,  say $x_1, \ldots, x_n$, 
whose set of values constitute the ($n$-dimensional) {\em state-space}
of the  system. Any other parameter $y$ is then  given in terms of these by 
an {\em equation state} (or {\em constitutive equation}), 
$y = f(x_1, \ldots, x_n)$. In particular, the quintessentially thermal
parameter, {\em temperature}, characterizes equilibrium, which is 
the content of
\begin{itemize}
\item{The Zeroth Law}: a state of equilibrium exists; equality of temperature
is a necessary condition for thermal equilibrium between two systems.
\end{itemize}
 
The simplest  example of thermodynamic system is that of a
one-component chemically inert homogeneous fluid (liquid or gas) 
in a container of volume $V$ at temperature $T$. Its state 
space could be taken as the two-dimensional set of points, say $(V,T)$, in 
the first quadrant. All other state variables can be obtained as functions 
of $(V,T)$ through the equation of state, for instance the pressure  
$p = f(V,T)$. For example,  for an  ideal gas one has
$\displaystyle p = N\, k \, T / V$, where $N$ is the number of molecules
and $k$ is Boltzmann's constant; for the (non ideal) van der Waals gas
$p = N\, k\, T/(V- b) - a/V^2$ (with suitable constants $a$ and $b$).

The fundamental problem of classical thermodynamics might then be 
formulated as follows: given an isolated system in an initial 
equilibrium state, find the final equilibrium state to which the system 
relaxes, after some internal constraint had been lifted. Here, there is
an implicit {\em dynamical}  assumption (experimentally supported), 
namely, that an isolated system, when left to itself, will eventually 
reach (``relax to'')  an  equilibrium state: this is the
{\em trend to equilibrium} property. However, as there is as yet no reference 
whatsoever to a time parameter,  the mention of  dynamics at this stage 
seems to have only a motivational or heuristic purpose. In other words, 
classical thermodynamics  would be concerned only with the {\em outcome} of 
the potentially very complex and  violent happenings which the system 
experiences in its (time) evolution between the initial and final 
equilibrium states.

In any event,  the First Law of Thermodynamics (Conservation of Energy) is then
stated and taken to hold for any kind of thermodynamic ``transformation'' 
or ``process'':
\begin{itemize}
\item{The First Law}: To every thermodynamic system there is associated
a state variable, its internal energy $U$, such that
in every infinitesimal transformation (``process''),
$$
dU = dQ - d W,
$$
where  $dQ$ is the heat absorbed by the system and  $dW$ the worked performed
by it (in particular,  in an isolated system the internal energy is 
conserved).  
\end{itemize}
Sometimes this is said to provide a definition
of heat in terms of work, but if so, we would not be dealing with 
a law of nature but just a definition!~\footnote{A similar mistake is
sometimes made in some textbook presentation of newtonian mechanics, where 
Newton's second  law is  said to provide definition of force, which is in 
fact a primitive concept there.} In the usual formulations of thermodynamics, 
heat is a primitive concept, its inter-convertibility  into work and internal 
energy being the crucial aspect of the first law.

 While at this stage, a reference to ``transformations'' still does not cause 
much harm, things get increasingly confusing in the formulation of the Second 
Law,  where the notions of reversible and irreversible {\em processes}  
explicitly  appear. 
\begin{itemize}
\item{Second Law of Thermodynamics}: There is a state variable, the
{\em entropy} $S$, such that for  {\em reversible} processes 
(in non-isolated systems), $dS = dQ/T$, where $T$ is the absolute temperature;
in {\em isolated} systems, for {\em irreversible} processes, 
the entropy never decreases.
\end{itemize}

A dynamical aspect of the theory apparently enters the picture the moment 
the notion of ``process'' is mentioned. The trouble again is that, 
while by  {\em process} one usually means a change 
of states {\em in time}, there is no explicit time parameter in the previous
discussion: after all, one is dealing only with equilibrium states, which
are supposedly time-independent. Besides, in  real systems, for instance
fluids, a change from an equilibrium state to another inevitably involves
some (at least local) space and time inhomogeneity; therefore the
basic quantities describing the system become time-dependent {\em fields}, 
so that during the process, the state space of the system is no longer a 
finite-dimensional manifold as before, but an infinite-dimensional one.

It is also not quite  clear what is meant by a reversible process. 
In principle, it is a process that could be undone, that is, to which 
there is associated  another process  consisting in the reversed order of 
states {\em in time}.  It seems that classical equilibrium thermodynamics 
deals  only with these  kind of processes which, on the other hand, are 
sometimes said not to be, strictly speaking,  processes at  all, but  just
``sequences of states  of  equilibrium''.~\cite{Somme}  Also, one usually 
depicts such ``processes'' as paths in state-space, supposed sufficiently 
smooth so that  some path-integrals can be performed, and  it would  seem 
natural that these paths should be parametrized by time! 

A common way out of this confusion is to say that reversible processes are 
only idealizations of real processes, which are always irreversible 
(in particular, not representable in general as smooth paths in state space).  
For heuristic purposes, so goes the argument, one can consider
this idealization as a good approximation to real (time-dependent) processes 
in the limit of zero rates. These so-called ``quasi-static processes'' are 
conceived of as evolving through ``infinitely slow and sufficiently small 
steps'' in such a way that at each instant the system immediately relaxes to 
an equilibrium state. They are not only  heuristic devices,  however, but are 
crucial  calculation tools. For example, to calculate the  entropy 
change between two equilibrium states one {\em imagines} a reversible 
process connecting those same two states. But the 
feasibility, in principle, of this procedure, is rarely discussed: should it 
not either be proven or clearly taken as a hypothesis for each thermodynamic 
system (say,  as  a ``state-accessibility'' property)? In any case
such notions are very rarely treated with the care they 
deserve.~\footnote{A  rare example of a  clear-cut and mathematically 
precise  treatment  of such ``quasi-static 
processes'' (of course in the context of time-dependent changes of state, 
i.e., processes properly speaking) can be found in Ref. 38} 

It therefore seems that,  as is the case with  mechanics, a distinction 
should be clearly made between two branches of thermodynamics: classical 
equilibrium thermodynamics, which is really thermo{\em statics},  
concerned only  with  {\em equilibrium} states and their properties 
(like stability, etc) and  where time plays no fundamental role; this is what
textbooks' discussions of the three laws probably refer to. And  general 
{\em non-equilibrium}  thermodynamics, dealing with time-dependent 
phenomena  including, but going beyond, equilibrium 
states~\footnote{Which, by the way,  should  be obtained  as special states, 
not only  {\em stationary}   (i.e., time-independent) but also such  that 
temperature is uniform throughout the system~\cite{Syl}.} and explicitly
involving  the concepts of time, processes and  dynamics. That this is a 
much more complicated and less developed branch, and whether there 
is (or there could be) a unified treatment of it, are extremely important 
but separate issues.

 Now, for a simple fluid, the First Law joined to the first part of 
the Second Law implies that for infinitesimal reversible processes the 
{\em fundamental equation}  of equilibrium thermodynamics for homogeneous
fluids (or Gibbs relation) reads:
$$
dS = \frac{dU + p \, dV}{T}.
$$

One of the tasks of equilibrium statistical mechanics would be to somehow
derive this fundamental macroscopic relation from microscopic principles. On 
the  other hand, to study  transport phenomena such as diffusion, viscous 
flow, conductivity, and also to (hopefully) elucidate  the 
trend to equilibrium issue, one needs to enter the realm of 
out-of-equilibrium systems.

In sum, thermodynamics is an incredibly successful theory, in spite
of having been  marred by a long history of conceptual problems. It is an 
interesting, rich an live theory with many open problems. Still, it is a 
phenomenological  theory (or of black-box kind) in the sense that there is no 
hint about the underlying mechanism that  could explain the thermodynamic 
laws in terms of more basic  (i.e., microscopic) constituents. The aim 
(or should one say dream?) of statistical mechanics is to provide a unified 
microscopic explanation of  equilibrium and  non-equilibrium thermodynamics. 
This leads us to the next ingredients in the formation of statistical 
mechanics.

\subsection{Atomism, Mechanics, Kinetic Theory and Probability}

 Of the these ingredients, atomism was an ancient philosophical doctrine, 
while mechanics came to age at the scientific revolution, having attained
its zenith in the  developments of analytical mechanics during the
mid-XIX$^{\mbox {\scriptsize th}}$ century. As for kinetic 
theory, it is a kind of blending of these two previous ingredients 
plus the somewhat surprising role of probability, with the aim of 
providing a mechanical-atomistic explanation of the behavior of
gases. Let us briefly discuss these contributions.

The atomic theory of matter, or {\em atomism }, is one of the most
daring and  fruitful ideas of the early greek 
philosophers.~\footnote{Particularly associated to  Democritus of Abdera,
fifth century B.C. and also to some ancient Hindu sources.} Though totally 
speculative and  qualitative in its origins, it turned out to be 
(at least in general lines) the accepted  viewpoint of modern physics.  
Of course, we can only say that  with the hindsight of 2500 years of 
enduring controversy and painstaking research. And, in fact, the actual 
atomic structure of matter is much more complicated than  could have ever 
been conceived in the fifth century B.C.: first and 
foremost, atoms are not really indivisible, having a complex internal 
structure, the understanding of which  demands mastering the sophisticated 
mathematical and conceptual apparatus of quantum mechanics and relativity
theory.

In our ``post-atomic'' era,  in which  atoms can be photographed using 
electron-tunneling microscopes and even  manipulated individually with 
the help of laser tweezers, their reality is an almost banal fact.  Even so, 
it should not prevent us from  appreciating the boldness and innovation of 
atomism.~\footnote{In R. P. Feynman's eloquent words~\cite{Feyn}: ``If, in 
some cataclysm, all of scientific knowledge  were to be destroyed, and only 
one sentence passed on to the next generation of creatures, what statement 
would contain the most information in the fewest words? I believe it is the 
{\em atomic hypothesis}...or {\em atomic fact}...''} The very notion that 
observable properties of things could be explained through the complex 
arrangements of some hypothetical (invisible) discrete material entities was 
extremely controversial (to begin with, it was 
quite counterintuitive).~\footnote{Today, however, we  recognize the 
procedure of postulating the existence of some 
{\em material} invisible entities in order to explain complex 
phenomena, as one of the hall-marks of modern science. Of course, 
with the crucial proviso that the hypothesized entities should not 
be inscrutable, having in each case to be subjected to
careful experimental (even if very indirect) testability.}

It is therefore not surprising that very soon after its proposal, the
atomic theory had a rival, rather commonsensical, {\em continuum theory} 
(a byproduct of the stoic school), according to which 
the continuous substances provided the foundations for all natural phenomena, 
without the need of invoking invisible entities.~\footnote{This 
idea would find its modern counterpart in the various field theories
of physics, like continuum mechanics, hydrodynamics, electromagnetism, etc. 
Incidentally, as in any deep theory, these ones contain plenty of
``unobservables''.~\cite{Bun}} We can already discern here the seeds of the 
future quarrel between the atomists and the so-called
``energeticists'' in  the last half of the 
XIX$^{\mbox {\scriptsize th}}$ century, over the {\em existence}  of 
atoms.~\cite{Cer}  That controversy happened in the context of the then 
new {\em kinetic theory of gases}, greatly advanced by Maxwell and Boltzmann.

Kinetic theory is an attempt to use the atomic theory of matter
and mechanics to explain the thermodynamic  behavior of gases, 
being an early reductionistic program of physics.~\footnote{Note the 
prominent  status and role of mechanics, even at a time when the field 
theories of physics, in particular electromagnetism, were gaining acceptance. 
The weight  of the mechanistic viewpoint is clearly seen by the fact that 
Maxwell himself tried to interpret the electromagnetic fields as mechanical 
vibrations of an hypothetical  ether.} Starting with the pioneering
paper by Clausius entitled ``The kind of motion we call heat'' (1857),
gases were pictured as being made of a huge  number  microscopic particles 
or molecules (of the order of $2.7 \cdot 10^{19}$ per cubic centimeter at 
1 atm and $0^{\circ}$C). In the simplest model, the particles are taken as 
tiny rigid balls (of size of the order $10^{-8}$cm), interacting according 
to the laws of classical mechanics, namely, through  elastic collisions. 
These collisions would somehow provide the basis for an explanation of 
macroscopic phenomena; for instance, the pressure of a gas would be the  
result of the collisions of particles with the container walls. In this way 
one would ultimately be able to  explain the laws of thermodynamics, 
providing a  ``the mechanical theory of heat''.~\cite{Brush} 

This program had some startling initial successes in the work of Maxwell 
(for example,  his prediction that fluid viscosity is independent of 
density, for low-density fluids). It was further developed by Boltzmann, 
amid a growing resistance from the anti-atomists.~\footnote{As mentioned 
before, one has to remember that at
that time the existence of atoms was far from being universally accepted. 
It was Einstein's 1905 work on Brownian motion (using statistical mechanical 
ideas!) which finally settled the issue.} 
Particularly important was the proposal of the Maxwell-Boltzmann transport
equation describing the  time evolution of the distribution function 
$f({\bf r}, {\bf v}, t)$, where $f({\bf r}, {\bf v}, t) \, d^3 {\bf r} \,
d^3{\bf v}$ is interpreted as the number of gas particles in the volume
$d^3 {\bf r} \, d^3{\bf v}$ around ${\bf r}$ and
${\bf v}$ at the time $t$. Namely:
$$
\frac{\partial f({\bf r}, {\bf v}, t)}{\partial t} + 
{\bf v} . \nabla f({\bf r}, {\bf v}, t) = Q(f,f), 
$$
where the right-hand term (the so-called collision term) summarizes  the 
effects of collisions. 

This is probably the very first (integro-) differential 
equation for the time-evolution of  a {\em probability density} 
(after normalization).  This  equation 
was ``deduced'' by Boltzmann, for the case of dilute gases, from heuristic 
considerations  of binary particle collisions, plus some additional  
hypothesis on the initial conditions (the  famous  ``molecular chaos 
hypothesis''). From this equation Boltzmann obtained his startling 
``H-theorem'', which seemed to provide for the first time a 
derivation of the relaxation of a gas to equilibrium. This, however, attracted
sharp criticisms and  generated a lot of controversy, particularly in 
connection to the so-called ``irreversibility
problem/paradox''.~\footnote{In the ensuing debate, among other things, 
Boltzmann proposed his famous ergodic hypothesis.}

Without entering into a detailed discussion of such issues,~\cite{Cer} to 
which we intend to return in another occasion (in the the context of 
non-equilibrium problems), we observe a very important novelty: the 
introduction (others would say intrusion) of {\em probabilistic considerations 
into mechanical problems.}  

One should bear in mind that, although probability was by then a somewhat 
familiar topic, it nonetheless had  a very confusing status. Some people 
thought it was part of physics,  others that it just consisted of some set 
of guiding rules  for ``reasoning under uncertainty'' or gambling, and yet 
others thought that it provided general principles for organizing large 
chunks of data (with the emergence of the fields of 
statistics, insurance and demography).  

Probabilistic concepts had undergone great developments  since its beginnings 
in  1654, in the famous correspondence of Pascal and  Fermat on the division 
of stakes in  games of chance. A great impetus came from the need to 
understand the statistical regularities observed in certain  ``random'' 
phenomena involving  a large number of trials (or repetitions) of 
similar occurrences. For example, the stabilization of the relative frequency
of heads in coin-tossing games (a manifestation of the Law of Large Numbers)
and the ubiquity of the normal 
(or Gaussian) distribution (connected to
the Central Limit Theorem), ranging from the errors in 
astronomical measurements through the height of conscripts in the 
military.~\footnote{Interestingly, the discovery of 
such statistical regularities in social affairs, such as demography,
seemed to  corroborate Adolphe Quetelet's program of a ``social physics'', 
and, apparently, these ideas percolated into the physical sciences, being
one of the few occasions when the mutual  influence was in this 
direction.~\cite{vonP}}. However, probability was not as yet a 
theory proper, but rather a collection of more or less general results. 

It was only in 1933 that it finally reached  maturity with
the axiomatization provided by A. N. Kolmogorov~\footnote{There was some
previous proposals, but none has got such immediate and universal
acceptance from the mathematical community as Kolmogorov's.} in his classical 
treatise,~\cite{Kolmo} which greatly helped in clarifying its nature. In 
the first place it became clear, once and for all, that probability theory, 
like geometry and analysis, is a branch of pure 
mathematics, not  of physics. As such, it has many possible 
models (in the set-theoretic sense, i.e., examples or realizations in
mathematics) and many different interpretations in applications to the
factual sciences.~\cite{Bun3} In particular, one need not be {\em ab initio} 
committed to any given interpretation,  be it subjectivistic 
(as degrees of belief), frequentist  (stabilization of frequencies of 
repeated trials), the  propensity view or any other. As a matter 
of fact, once the formal structure of the theory have been  elucidated, the 
adequacy of any suggested interpretation, vis-a-vis some intended application,
could be better examined, criticized and justified.

The great insight of Kolmogorov was to notice that, besides the 
standard ``elementary'' probability theory, that is, that part dealing 
with discrete arrangements of many objects (usually under the hypothesis
of equal probability) and which essentially reduces to (usually very
intricate) combinatorics, there is a more general part which included  
some well-known classical cases involving so-called continuous 
distributions. He noticed that the adequate unifying framework 
would be provided by the then recently created 
{\em measure theory}.~\cite{Bing}  That is the theory proposed in Henri 
Lebesgue's 1905 doctorate  thesis, which is a generalization of the concepts 
of length, area and volume.~\footnote{This theory  is the culmination
of some internal developments in classical mathematical analysis, linked to 
the clarification of the notion of function, Fourier series and integration 
theory. In particular it gave  an extension of the Riemann integral, having
many desirable properties. Specifically, it  allows,
under very general conditions, to take limits inside the integral sign,
for sequences of functions, as in the classical monotone convergence theorem
and dominated convergence theorem. In its general abstract version, measure 
theory strongly influenced  virtually all branches of mathematics.} 

We next describe the main ideas in the precise formulation of the 
statistical mechanics program.

\noindent\section{Equilibrium Statistical Mechanics} 

Statistical mechanics main aim is to deduce the ``collective'', ``emergent''
or ``macroscopic'' behavior of a system composed of a large number of 
microscopic interacting particles. We note that there is nothing mysterious 
regarding  emergent properties: these are just properties of the system 
which the individual components lack, e.g., temperature for a particle
system.

The main idea is that, in equilibrium, the microscopic dynamical
details are not important or relevant, and the macroscopic properties
appear as certain {\em averages}  with respect to a suitable family of
probability measures on phase-space: the so-called {\em ensembles}. 
Here, a crucial link with statistics is the fact that one is dealing with 
systems consisting of an extremely large number of microscopic components.

\subsection{The Microscopic Model}

In classical statistical mechanics the microscopic 
model of a fluid in a container consists of  $N$ identical and 
structureless (point) particles with mass $m$,  located in  a  
subset  $\Lambda \in {\RR}^3$ and evolving according to the  laws of 
classical mechanics.~\footnote{In the  somewhat misleading jargon of 
statistical mechanics, these models are referred to as ``continuous''  models, 
as they  allow particles to move in the space continuum $\RR^3$, in contrast 
with  `` discrete'' lattice-gas models, in which particles can
only occupy the discrete sites of a lattice. Of course, both are discrete
models of the microworld, in line with the atomistic viewpoint.} 

Though admittedly a caricature of microphysics,  this model is still 
more realistic than the one provided by {\em lattice models}, at least for 
fluids. In fact, lattice systems  are highly idealized pictures  of  
microphysics, more appropriate for describing crystalline 
systems, where the atomic  motions are so restricted that it is a good 
approximation to suppose that they can only occupy the sites of a lattice. 
Moreover, in contrast to the  Hamiltonian dynamics of classical
mechanical particles, lattice systems don't have a natural  
dynamics, which  is usually  imposed in an {\em ad hoc} fashion (and usually 
taken to be intrinsically stochastic).~\footnote{This would not be 
too problematic, however, as  long as one is dealing with equilibrium 
statistical mechanics which, as we will see, ignores the details of dynamics.
This seems to justify some kind of ``model-independence'' of the results
of statistical mechanics which in turn would further justify the study
of idealized models.}

That said, one has to recognize that most of our detailed  knowledge of
statistical mechanics comes from the study  of lattice systems, which is
one of the greatest achievements  of modern mathematical-physics. It is 
a huge research field, with a long  history of successes, based
on a  rigorous analysis of diverse idealized models.
Moreover, it is a fundamental source (as well as a test field) of a 
variety of ideas and concepts which are at the core of  our understanding
of statistical mechanics.~\cite{Minlos, Geor, vanEnt} 

Ultimately, of course, a physically realistic model should begin from a 
quantum mechanical formulation (say, non-relativistic) for the basic 
atomic-molecular model. However, for historical reasons 
(i.e., kinetic theory) some of the first rigorous results were 
achieved within the classical framework,  even within the rigid ball model.
Far from trivial, it is nonetheless somewhat simpler 
and surprisingly adequate.~\cite{Cohen}  As  J. Lebowitz 
remarked~\cite{Lebo}
\begin{quote}
 \textsf{Why this crude classical picture (a refined version of that held by some 
ancient Greek philosophers) gives predictions that are not only qualitatively
but in many cases also highly accurate, is certainly far from clear to me...}
\end{quote}

In the chosen model, the {\em microstate} of the system consists of 
the positions and momenta of all particles, that is, of  a point 
$\displaystyle  \omega = ({\sf q},{\sf  p}) = ({\bf q}_1,  {\bf p}_1, \dots,
 {\bf q}_N,  {\bf p}_N)$ in the system's {\em phase-space} (or state-space)  
$\Omega_{N,\Lambda} = (\Lambda \times {\RR}^d)^N$.

Suppose, for simplicity, that  $\Lambda = \RR^3$. The  time-evolution 
(or dynamics) of the system is given by Hamilton's equations:
\begin{equation}
\displaystyle\left\{
\begin{array}{ll}
\displaystyle\frac{d{\bf q}_i(t)}{dt}  = - 
\frac{\partial H({\sf q}(t), {\sf p}(t) )}{\partial {\bf p}_i (t)}  & \\
& \\
\displaystyle\frac{d{\bf p}_i(t)}{dt}  = - \frac{\partial H( 
{\sf q}(t),{\sf p}(t))}{\partial {\bf q}_i (t)} & ,
\end{array}
\right.
\end{equation}
{\em  plus} the initial data 
$({\sf q}(0), {\sf p}(0)) = ({\sf q}_0, {\sf p}_0)$ (for convenience, we took
$t_0 = 0$).~\footnote{That
the initial data are an {\em integral part}  of the dynamical 
description of a mechanical system, though a  trivial observation,
is useful bearing in mind, particularly  regarding the question of
reversibility in kinetic theory.}

Here, the  {\em Hamiltonian} (or total energy)  
$H(\omega) = H_{N,\Lambda}(\omega)$  of the system is a real-valued function 
on phase-space given by 
$$
H({\sf q},{\sf p}) = \sum_{i=1}^N \frac{{\bf p}_i^2}{2m} + 
\sum_{i < j} \varphi(|{\bf q}_i - {\bf q}_j|),
$$ 
where $m > 0$ is the mass of each particle
and $\varphi(\cdot)$ is a central pair-potential interaction 
energy.~\footnote{We will consider only this class of {\em separable} 
Hamiltonians, that is, for which 
the momenta and position variables are segregated in different terms. More
general non-separable Hamiltonians can be very important; 
for example in the two-dimensional vortex model in  fluid dynamics one  
deals with the non-separable Hamiltonian 
$\displaystyle H({\sf p, q}) = -\frac{1}{8\pi} \sum_{i,j=1, i\neq j}^{N}
a_i \, a_j \ln[({\bf q}_i -{\bf q}_j)^2 + (\frac{{\bf p}_i}{a_i}-
\frac{{\bf p}_j}{a_j})^2]$, where
the $a_j$'s are some parameters.}

If $\varphi$ is sufficiently smooth (say, twice continuously differentiable),
and short-ranged, then standard ordinary differential equations  theory 
guarantees the existence and uniqueness of {\em local} solutions. That is, 
functions ${\sf p}(t) = 
{\sf p}({\sf q}_0,{\sf p}_0;t)$,
${\sf q}(t) = {\sf q}({\sf q}_0,{\sf p}_0;t)$, defined for some finite open
time interval $a < t < b$,  
which are differentiable functions of the initial data 
$({\sf q}_0, {\sf p}_0)$ 
and of time, satisfying  equations (1).
Moreover, the solution can 
be extended to a global one, i.e., for $-\infty < t < +\infty$. It
thus defines a trajectory or {\em orbit} (i.e., a smooth curve) in phase-space.

So, for each $t \in \RR$ one defines a  dynamical {\em flow} $T_t$, 
taking each initial data $({\sf q}, {\sf p})$ to its $t$-evolved image 
under the dynamics,
\begin{equation}
\begin{array}{ll} 
T_t: \RR^{3N} \times  \RR^{3N} \longmapsto \RR^{3N} \times  \RR^{3N}&\\
({\sf q},{\sf  p}) \longrightarrow ({\sf q}(t),{\sf p}(t)) = 
T_t({\sf q},{\sf p})&, 
\end{array}
\end{equation}
the set $\{T_t \, : \, t \in \RR\}$ being  a one-parameter group of 
transformations,
i.e.
\begin{equation}
\left\{
\begin{array}{ll}
T_0 = {\bf 1} & \\
T_t .  T_s = T_{t+s} & \\
T_t^{-1} = T_{-t} &.
\end{array}
\right.
\end{equation}

As is well known, Hamiltonian flows (even local ones) have the following
two fundamental properties:
\begin{itemize}
\item[1.] Energy is an integral of motion: for all $t$,
$$H(T_t({\sf q},{\sf p})) = H({\sf q},{\sf p});$$
\item[2.] {\em Liouville's theorem}: Lebesgue measure (volume) 
$\lambda_N$ on phase-space is  invariant, i.e., for every 
measurable set $A$, and for all $t$
$$
\lambda_N(T_t^{-1}A) = \lambda_N (A),
$$
where
$$
\lambda_N(A) \equiv \int_A \Pi_{i=1}^N d^3{\bf q}_i \,  d^3{\bf p}_i . 
$$
\end{itemize}

Liouville's theorem is an extremely important fact: it 
says that there is a {\em natural} invariant measure around, 
namely Lebesgue measure on phase space, which is crucial to the ensemble 
theory. Energy conservation implies that the orbits are restricted to the 
energy surface defined  by $H({\sf q}, {\sf p}) = E$, where $E$ is the 
initial energy of the  
system.~\footnote{If there are additional conserved quantities, the
motion is of course restricted to the intersection of
the corresponding surfaces. We observe that if the energy surface is a compact
set the existence of an invariant measure for the dynamics follows from
{\em Krylov-Bogolyubov's theorem}.}

The basic  dynamical issues can be more involved in the case of singular 
potentials (e.g., in celestial mechanics), where even global existence
of the flow is not warranted due, for example, to so-called 
collision singularities. However, for gases one typically works with the 
Lennard-Jones potential, a semi-empirical potential of the form
$$
\varphi(r) = \varphi_0 \Big [ \big (\frac{r_0}{r}\big )^{12} -
\big (\frac{r_0}{r}\big)^6\Big],
$$ 
with strength $\varphi_0$ ($r_0$ is the point of minimum of the potential). 
This is a popular choice of potential giving  a qualitatively
realistic  description of molecular interaction for inert gases: strong short 
range repulsion and weak long range attraction. Being bounded from below,
there is no catasthropic collision singularities. Alternatively, one can work
with hard-spheres which move freely and interact only 
through elastic collisions. An  additional 
complication is the  confinement issue, namely that particles are supposed to 
be restricted to a bounded region  (container) $\Lambda \subset \RR^3$.

Though a bit harder to establish, the main properties of
the flow can be obtained for  those cases also.
The details, though  very  important for the {\em dynamical}  foundations of 
statistical mechanics, are not so relevant to  the ensemble theory of 
equilibrium  statistical mechanics, which is the focus of this paper.
As we will see, in this  context the dynamics is, so to speak, swept  
under the rug, once  the ensembles are  identified  to certain 
{\em invariant} probability measures on  phase-space.

\subsection{The ensembles}

One might at first get the impression that there is 
a kind of built-in duality in the foundations  of classical statistical 
mechanics, reflected in its very name, which juxtaposes two apparently 
antithetical concepts: mechanics and probability (or statistics). 
That is, though starting from a microscopic system of interacting 
newtonian particles, there soon appears, as if by fiat,  a statistical or 
probabilistic ingredient, which is supposedly alien from the classical world. 

The justification of that situation begins with the standard
operational argument: it is impossible to know the microstate of
such huge particle systems (as  one cannot, in practice,
simultaneously measure each and every particle's position and momentum); 
moreover, so the argument goes, even if the microstate were accurately known, 
it would be hopeless to solve a system of the order of $10^{23}$ differential
equations. In sum, one has to use other means to study such systems
and that is where statistics comes to the rescue.~\footnote{This kind of argument seems to have 
been borrowed from the highly influential operational philosophy of standard 
quantum mechanics. It is  also to blame for conveying the misleading idea 
that the microstate of the (classical) system is a probability measure 
instead of a point in phase space.} 

Although it has a grain of truth, this rationale is somewhat confusing
and has to be qualified in many respects. First of all, it mixes
theoretical, epistemological and even methodological concepts, which should 
be kept separated.  For example, {\em our} inability to measure  
the initial data with infinite precision is certainly an unavoidable fact,
having very important methodological consequences bearing on the experimental
analysis of models and the limits on predictability (for example, 
in meteorological systems and chaotic dynamical systems).
However, such issues do not refer to the physical system  
the equations are supposed to model, which doesn't care about
human limitations. Besides, imprecision in measurement happens even for 
systems of few particles, so it is not intrinsically linked to the large 
numbers involved in statistical mechanics. 

As for the  ``solvability'' issue of the dynamical equations
(although not that important for equilibrium
statistical mechanics), similar  observations could be made:
the solvability of equations is an important {\em mathematical} 
(not physical)  question. But in order to state it 
correctly, one has to carefully and rigorously explain what it means  to 
solve or ``integrate''  a certain system of differential equations 
(for example, a series solution qualifies or not?).  Once in possession of 
such a notion and also of a way to survey the collection  of all 
differential equations  of a given kind (e.g., with the aid of
a topological notion of size), one can then proceed to  examine
whether ``most'' of the equations  are solvable,  or whether a particular 
one is.~\footnote{An illuminating example is the three-body
problem in celestial mechanics: it is non-integrable 
(i.e., cannot be algebraically solved), though it has a 
convergent series solution (hence an analytic solution) whose rate of
convergence is too slow to be useful to understand the long-time behavior
of the system!~\cite{Diacu}}

Furthermore, the claim that it is hopeless to solve a huge system of equations 
is not correct in all generality and depends on the integrability 
properties of the system. So, for example, a Hamiltonian system consisting 
of an arbitrary number of harmonic oscillators is perfectly solvable and one 
can write down the solutions explicitly.~\footnote{Another, less trivial, 
example is the Toda lattice system which, though highly non-linear, is 
completely integrable.}

It is frequently stated that while microscopic systems are
very ``complex'' (by which it is usually  meant having a
great number of degrees of freedom), macroscopic systems are much simpler, 
being described by very few variables and equations. This drastic 
``decimation'' of degrees of freedom, characterizing the  passage from the 
microscopic to the  macroscopic description, suggests the use of an 
{\em averaging  procedure},  and hence of  statistics. This viewpoint is much 
more sensible, focusing as it does on the role of statistics as a 
level-bridging ingredient, connecting the micro and macro realities.

We remark, however,  that while  {\em some} macroscopic systems (for
example, homogeneous fluids) do  have a relatively simple description 
{\em in equilibrium}, they  can be  extremely  complicated in the 
{\em non-equilibrium} case, as testified  by the (poorly understood) 
phenomena of turbulence. There, the  motion is described by time-dependent 
{\em fields},  that is, {\em infinite-dimensional}  
vectors,~\footnote{Note also that similar qualms could be raised here 
regarding  ``practical measurability'' of the precise state of the fluid:
the situation is even worse because fields, being an infinite component 
vectors, cannot be  measured completely not only in practice but in 
principle. However this never  prevented the study of fluid dynamics.} 
so that the  decimation mentioned  above is 
illusory. Moreover, such fields satisfy certain non-linear partial 
differential  equations which are, at present, beyond mathematical 
tractability.~\footnote{See, for example,
the Clay Mathematical Institute's million dollars prize for a proof of 
existence and smoothness for the  Navier-Stokes  equation.}

\subsubsection{The Boltzmann-Gibbs Principle}

It was Boltzmann who gave the clearest view of the situation of
statistical mechanics,  while struggling to answer the criticisms of
his results on  kinetic theory. His insight begins with the 
following simple but crucial observation:~\cite{Lebo}
let $F$ be a ``physically relevant'' state-function, that is,  a
function $F \,: \, \Omega_{\Lambda, N} \rightarrow \RR$  on phase-space to 
which there is a corresponding  macroscopic variable (typical examples are 
the ones associated with the conservation laws, like energy and momentum). 
Let ${\sf F}$ be a given  {\em equilibrium value} of that  macroscopic 
variable. Now, there are usually  very many different microscopic states 
$\omega \in \Omega_{\Lambda, N}$ compatible with the given macroscopic 
value.  For example, there are many different 
microstates  associated to the {\em same} value of total energy. 
It then makes sense to consider the subset 
$\Gamma_{\sf F} = \{ \omega \in \Omega_{\Lambda, N}
\, : \, {\sf F}=  F(\omega) \}$ of phase-space, consisting of all 
those microstates, as they are the ones putatively relevant to the 
micro-macro change of description. 

It is then quite natural to ask oneself about  the relative ``sizes'' of 
such  subsets with respect to the whole phase-space, in order, for example, to 
assess their ``relevance''  as compared to any other subset. One possible 
notion of size is the relative volume in phase space, as defined by the
Lebesgue measure which, by Liouville's theorem, is
invariant under the dynamics. In this way one  focuses in the ``fraction'' 
of states in phase-space corresponding  (or relevant) to the given value of
the  associated macrovariable.  This amounts to nothing more than 
``counting'' phase-space points,  that is, a sort of 
(continuous) ``combinatorial'' estimate of certain subsets, using  relative 
volume as the yardstick. 

As such, there is no  ``chance mechanism'' involved here, no more than
when comparing volumes of geometrical figures. Nor is necessarily involved 
any notion of ``choosing states at random''  or of
``ignorance'' about the state of the system. Now, in  the case of a compact 
phase-space, its total volume being finite, one  can normalize the Lebesgue
measure and we end up with a {\em probability measure}  $\P$ on phase-space 
(or on the energy surface); hence all the relevant techniques and results of 
probability theory apply.

Boltzmann and Gibbs then made a bold hypothesis: they proposed as the
{\em fundamental postulate of equilibrium 
statistical mechanics} that, for any  physically 
relevant state-function 
$F \,: \, \Omega_{\Lambda, N} \rightarrow \RR$,
the corresponding  macroscopic equilibrium value is given by its 
expected (or mean) value) with respect to a suitable {\em invariant} 
probability measure $\P$ on phase-space, i.e.,
$$
{\sf F} = <F>_{\P} = \int_{\Omega_{\Lambda, N}} F(\omega) \, \P(d\omega),
$$
at least when the number of particles $N \rightarrow +\infty$ (more on that
later).

Each such $\P$ is a member of a so-called {\em  ensemble}.  We emphasize 
that the procedure of taking averages~\footnote{Notice that, though  in  
probability theory one usually begins with a  probability measure and 
{\em then} proceeds to define
the expectation or average,  one could  take the opposite path; that is
(in case the sample space is  compact Haudsdorff space), beginning
with a non-negative linear functional $<\cdot>$ on continuous functions, it
can be  proved that there is a probability measure that represents this 
functional: this is the 
{\em Riesz-Markov representation theorem}.~\cite{Reed}} is not 
{\em necessarily} linked to  any random mechanism: it might  just mean
that details are unimportant.~\cite{Bun2} 

Of course, such a principle requires many clarifications and raises 
many  questions. Which are the ``suitable'' probability measures 
and why? Are they unique? Which are the (class of) relevant state-variables? 
What does the limit $N \rightarrow \infty$ mean? 

Let us begin with some nomenclature. As we have seen, from the viewpoint
of modern mathematical-physics, an  {\em ensemble} is just
a family $\cE$ of invariant probability measures on phase space. 
More precisely, each  $\P \in \cE$ is indexed by some macroscopic 
(thermodynamic) parameters (e.g., volume, energy),  adequate to describe the 
physical situation of the (equilibrium) system under study.  An ensemble 
element is sometimes referred to as a  ``statistical state'' of the system,  
which probably means that such measures are to be identified with the 
macrostates of the system.  We submit that this is  misleading and should  
be avoided: as discussed  before, the microscopic state is a point of 
phase-space while the macroscopic state, for example, of a homogeneous fluid 
is, say, a pair of temperature and pressure values. So neither the macroscopic
nor the microscopic state are measures. So, what is the status of such 
measures? As each member of an ensemble refers to {\em both}  
the microscopic level (being a probability measure on phase-space) {\em and}
to  the macroscopic level (being indexed by the relevant macroscopic  
state-parameters), it can be  viewed as the  fundamental  
{\em level-linking} concept establishing  the connection of the  micro to 
the macro descriptions.
 
The requirement of invariance of the probability measures
seems quite natural when dealing with systems 
in equilibrium; and as will be apparent, in equilibrium statistical mechanics,
once an ensemble is chosen, the microscopic {\em dynamical} details are 
essentially forgotten in all the subsequent calculations of thermodynamic 
quantities. The microscopic {\em interactions}  are, of course, fundamental 
as will be testified by the crucial role played by the potential in the 
following. 

By Liouville's theorem, one  obvious choice of invariant measure is the 
Lebesgue measure (that is , volume) in  phase-space. But, of course one could 
ask why not choose another invariant measure, if any? And, more importantly, 
is there a microscopic {\em dynamical} justification of the Boltzmann-Gibbs
postulate? What would it be like? Those are perhaps the most difficult 
foundational questions of  statistical mechanics and which necessarily bear on
a deeper level of analysis, namely on non-equilibrium statistical mechanics. 
In spite of some advances, this is still
an essentially open question. Hence, a  more ``pragmatic'' justification
of the postulate (besides its coherence) is that it works fine in many 
physical applications, so that it is vindicated by its very success.

Concerning the actual form of the postulate, notice that besides the total 
particle number $N$ and total volume $V = |\Lambda|$,
some other physically relevant state-variables are:
\begin{itemize}
\item density (and specific volume): $\displaystyle\rho = 
\frac{N}{V} = \frac{1}{v}$;
\item total kinetic energy: $\displaystyle {\mathcal K}(\omega) = 
\sum_{i=1}^N \frac{{\bf p}_i^2}{2m}$;
\item total potential energy: $\Phi(\omega) = \sum_{i < j} 
\varphi(|{\bf q}_i - {\bf q}_j|)$;
\item total energy: $\displaystyle H(\omega) =  {\mathcal K}(\omega) +
\Phi(\omega)$;
\item momentum change (impulse) per unit time and per unit surface area
transfered to container walls by collisions of particles when
in state $\omega$: $ {\mathcal P}(\omega)$.

\end{itemize}
So, according the Boltzmann-Gibbs postulate, for a given $\P \in \cE$, 
the corresponding  macroscopic variables (at the parameter values associated 
to  $\P$) are given by the mean values,
\begin{itemize}
\item mean density: $ \displaystyle \rho = < \rho>_{\P} = 
\int_{\Omega_{\Lambda,N}} \, \rho \, 
\P(d\omega) = \frac{N}{V}$;
\item mean kinetic energy $\displaystyle K  = 
<{\mathcal K}>_{\P} = \int_{\Omega_{\Lambda,N}} {\mathcal K}(\omega) \, 
\P(d\omega)$;
\item mean potential energy $\displaystyle \Phi  = <\Phi>_{\P} =
\int_{\Omega_{\Lambda,N}} \Phi(\omega) \, \P(d\omega)$;
\item mean total energy: $\displaystyle U = <H>_{\P} = 
\int_{\Omega_{\Lambda,N}} H(\omega) \, \P(d\omega)$;
\item mean pressure: $\displaystyle p =   <{\mathcal P}>_{\P} =
\int_{\Omega_{\Lambda,N}} {\mathcal P}(\omega) \, \P(d\omega)$. 
\end{itemize}
Note that these quantities are in general functions of $N$, $\Lambda$ and
other parameters indexing the ensemble measures.

A crucial property required of an ensemble 
is that it correctly describes the equilibrium thermodynamics 
of  the system. In the case of  homogeneous fluids, this can be made precise
by the following~\cite{Gall}
\begin{defi}
An ensemble is called {\em orthodic} if taking an infinitesimal change in
the parameters indexing each of its elements, the corresponding variations
of the  macroscopic variables $U$, $p$, $V$ and $T$ defined above, 
are such that 
$$
\frac{dU + p \, dV}{T}
$$
is an exact differential, at least when $N \rightarrow +\infty$,
$V \rightarrow \infty$ with $\displaystyle \frac{N}{V} \rightarrow constant$. 
Here  $\displaystyle T = \frac{2}{3 k} \, \kappa$, where $k$ is 
Boltzmann's and $\kappa$ the mean kinetic energy density.

\end{defi}
Orthodicity  is a natural requirement. In fact, for such an ensemble, the 
macroscopic variables can be identified to the familiar thermodynamic
variables satisfying the known thermodynamical relations; so in particular, the
absolute temperature $T$ would be  interpreted as average kinetic energy per
particle. Moreover,  orthodicity  guarantees that there is a function 
$S$ of the macroscopic state variables (say, of  $(p,V)$ or $(U,V)$), 
which can be interpreted as the thermodynamic entropy of the system. This
function is such that the fundamental equation of classical  equilibrium
thermodynamics (for homogeneous fluids), namely Gibbs relation,
$$
dS = \frac{dU + p \, dV}{T},
$$
is satisfied.

Summarizing, the fundamental postulate of equilibrium statistical
mechanics, the so-called {\em Boltzmann-Gibbs Principle}, is the claim 
that the equilibrium thermodynamics of a (simple fluid) 
system is described (in the sense just discussed) by an orthodic ensemble.

Let us  recall the three main  classes of ensembles: {\em microcanonical}, 
{\em  canonical} and {\em  grand-canonical}.

\subsubsection{The Microcanonical Ensemble}

The microcanonical ensemble is the one suitable for {\em isolated} systems. 
The phase-space is reduced to the {\em energy surface}: 
$\Omega_{\Lambda, N, U} = \{ \omega \in  \Omega_{\Lambda, N} \; : \; 
H (\omega) = U \}$, which  is a compact set (if the potential is 
bounded from below), invariant under the dynamics.

The corresponding invariant measure on $\Omega_{\Lambda, N, U}$ cannot 
simply be the full phase-space volume measure, because the energy surface 
(being a set of codimension one) has  Lebesgue measure zero. The alternative 
is to use the  ``Lebesgue measure cut to the  energy surface'',~\cite{Lan} 
defined as follows. 

First, let us assume  that the phase-space is 
``symmetrized'', that is, we identify any two microstates which differ by a 
permutation of particles (in other words, consider the identical particles to 
be indistinguishable). Then, if $\nabla H( \omega)$ is non-zero on the energy
surface, for any measurable set $A$ on the surface the following limit exists:
~\cite{Kinchin}
$$
\nu_{\Lambda, N, U}(A) \equiv \lim_{\Delta U \rightarrow 0} 
\frac{1}{\Delta U}
\int_{A \cap J_U} \, \frac{1}{N!}
\displaystyle d\lambda_N \, =\, 
\displaystyle\frac{1}{N!} \int_A \frac{d\sigma(x_U)}{\|\nabla H(x_U)\|},  
$$
where $\displaystyle J_U = \{\omega \in \Omega_{\Lambda, N} \; : \; 
U \leq H(\omega) \leq U + \Delta U\}$ and  $\sigma(\cdot)$ is the area measure 
on the energy surface. Moreover, being  a limit of invariant measures, the 
measure $\nu_{\Lambda, N, U}$ is also invariant (the factor $N!$ accounts for
the symmetrization of Lebesgue measure~\footnote{Strictly speaking,
let $\pi: (\Lambda \times \RR^3)^N \rightarrow \Omega_{\Lambda,N}$ be
the natural projection taking each ordered point $({\sf q}, {\sf p})$ to
the corresponding unordered one, namely
$\pi({\sf q}, {\sf p}) = \{{\sf q}, {\sf p}\}$. So, if 
$\lambda_N$ is the usual Lebesgue-measure on 
(the $\sigma$-algebra of) $(\Lambda \times \RR^3)^N$, the corresponding
symmetrized Lebesgue-measure $\bar{\lambda}$ 
on $\Omega_{\Lambda,N}$ is defined by
$ {\bar{\lambda}}_N(A) = \frac{1}{N!} \lambda_N(A)$, for any $A$ in
the corresponding $\sigma$-algebra ${\mathcal M}_{\Lambda,N}$. This is
usually shortened by writing $d{\bar{\lambda}}_N = \frac{1}{N!} d\lambda_N$.
Note that the Hamiltonian is symmetric under permutation so that
it is in fact a function of the unordered pair $\{{\sf q}, {\sf p}\}$.}).

Then, by definition,  the  {\em microcanonical ensemble} is the family 
of invariant probability measures 
$\displaystyle\P^{\mbox{\scriptsize mc}}_{\Lambda, N, U}$, 
parametrized by $\Lambda$, $N$ and $U$, such that, for any measurable set
$A \subset \Omega_{\Lambda, N, U}$,
$$
\P^{\mbox{\scriptsize mc}}_{\Lambda, N, U}(A) = 
\frac{\nu_{\Lambda, N, U}(A)}{Z_{\Lambda, N, U}},
$$
where the normalization factor 
$$
Z_{\Lambda, N, U} = \nu_{\Lambda, N, U}(\Omega_{\Lambda, N, U}),
$$
is called the microcanonical {\em partition function}. The partition
function  is just the total  $\nu$-measure  of the new phase-space 
$\Omega_{\Lambda, N, U}$, and it can be viewed as 
a (continuous) ``counting'' of all available microstates of the 
system.~\footnote{Its original german name is {\it Zustandsumme} 
or ``sum over states''.}

The microcanonical ensemble is orthodic in the {\em thermodynamic limit}  
which is a kind of ``infinite-volume limit'' of the system. At this stage, 
this limit appears to be a technical question only, and we will discuss some 
of its physical justifications in the next section. Let us, however,
describe the main aspects involved in its procedure.

First, one considers  an increasing and sufficiently regular space-filling 
sequence of regions~\footnote{Boxes will do, but very general shapes
are possible, as long as the rate of increase of surface area to volume 
ratio is suitably controlled.} 
$\{\Lambda_i\}_{i\geq 1}$, that is $\Lambda_i \subset \Lambda_{i+1}$ and  
$\displaystyle\cup_{i\geq1} \Lambda_i  = \RR^3$ (this is
indicated by writing $\Lambda \uparrow \RR^3$).
At the same time, let  $\{N_i\}_{i\geq1}$ and 
$\{U_i\}_{i\geq1}$  be increasing sequences of energies and particle numbers, 
respectively, such that $v_i = V_i/ N_i \rightarrow v = 1/\rho$ and $u_i =  U_i/N_i \rightarrow u$, as $i \uparrow \infty$. Then, the following limit 
exists:~\cite{Gall, Ruelle}
$$
s(u,v) =  \lim_{\Lambda \uparrow \RR^3, \frac{U}{N}\rightarrow u,
\frac{V}{N} \rightarrow v}  \frac{1}{N} k \, 
\ln Z_{\Lambda, N, U},
$$
where $k$ is Boltzmann's constant. 

Notice  Boltzmann's famous  formula for thermodynamic 
entropy as proportional to the logarithm of 
the ``number'' of microstates: $S(U,V) = k \ln Z_{\Lambda, N, U}$, so
$s(u,v)$ is naturally  interpreted as the entropy density (or specific entropy).

Moreover, the function $s(u,v)$ satisfies Gibbs' relation:
$$
ds = \frac{du + p \, dv}{T}.
$$
Here, $\displaystyle T = \frac{2}{3 k} \kappa$, where $\kappa$ is the limit 
microcanonical average kinetic energy density,
$$
\kappa(u,v) = \lim_{\Lambda \uparrow \RR^3, \frac{U}{N}\rightarrow u,
\frac{V}{N} \rightarrow v} 
<\frac{\mathcal K}{N}>^{\mbox{\scriptsize mc}}_{{\Lambda, N, U}} \, = \, 
\lim_{\Lambda \uparrow \RR^3, \frac{U}{N}\rightarrow u,
\frac{V}{N} \rightarrow v} <\frac{1}{N} \sum_{i=1}^N 
\frac{{\bf p}_i^2}{2m}>^{\mbox{\scriptsize mc}}_{{\Lambda, N, U}}.
$$
Note that this is a kind of (weak) ``law of large numbers'', as one is 
calculating an asymptotic (``large $N$'') limit of sums of random variables, 
in this case, the  particle's kinetic energy  
${\bf p}_i^2/2 m$.~\footnote{Unfortunately  the situation is much more 
complicated than the classical laws of large numbers,
which usually pressuposes independence. Here, due to various constraints on the
motion, one cannot expect the random variables to  be independent.}

We also have the limit average pressure,
$$
p = p (u,v) =  \lim_{\Lambda \uparrow \RR^3, \frac{U}{N}\rightarrow u,
\frac{V}{N} \rightarrow v}
<{\mathcal P}>^{\mbox{\scriptsize mc}}_{{\Lambda, N, U}}.
$$
So, if  $T = T(u,v)$ is interpreted as the absolute
temperature and $s(u,v)$ as the specific entropy, then 
(assuming differentiability) as  
$\displaystyle  ds = \frac{\partial s}{\partial u} \, du +
\frac{\partial s}{\partial v} \, dv$, it follows that 
$\displaystyle \frac{\partial s}{\partial u}(u,v)
= \frac{1}{T(u,v)}$ and $\displaystyle \frac{\partial s}{\partial v}(u,v)
= \frac{p(u,v)}{T(u,v)}$. By eliminating $u$ in these relation, one could 
obtain the equation of state of the fluid: $p = f(T,\rho)$ 
(in principle at least, though by no means a trivial task in 
practice~\cite{Gall, Griffs, Toda}).

We observe that there are two separate issues involved here: orthodicity and
the thermodynamic limit. It turns out that for
the microcanonical ensemble orthodicity only holds in the
thermodynamic limit,~\cite{Gall} which is then a prior issue. In fact,
the most  difficult part of the above  results is the proof of
the existence of the limit $s(u,v)$, in terms of  which the other limit 
quantities can be expressed. For this reason the question of existence of 
this limit is sometimes referred to as {\em the} problem of the thermodynamic 
limit {\em at the thermodynamical quantities level}. 

As would be expected, the existence proof of  such  limit will necessarily 
require some hypothesis on the interaction potential $\varphi(\cdot)$. 
We see here an interesting interplay (even if coming out of an
apparently purely technical issue), of the micro-macro change of description: 
for  the microcanonical ensemble  to provide the correct macroscopic 
description, one needs to impose some restrictions on possible types of
microscopic interactions. 

The restrictions typically are:
\smallskip

(a) {\em stability}: there is a constant $B >0$ such that in every space
configuration ${\sf q}= ({\bf q}_1, \ldots, {\bf q}_N)$ we have
$$
\Phi({\sf q}) = \sum_{i < j} 
\varphi(|{\bf q}_i - {\bf q}_j|) \geq - B \, N;
$$

(b) {\em temperedness}: there are constants $C > 0$, $R>0$ and $x >0$
such that
$$
\varphi(|{\bf q}_i - {\bf q}_j|) \leq \frac{C}{|{\bf q}_i - {\bf q}_j|^{3+x}},
\;\;\;\; \mbox{for} \;\;\;\; |{\bf q}_i - {\bf q}_j| > R.
$$

These requirements are designed so that the thermodynamic limit exists.  
The stability condition  avoids a possible collapse of the 
system~\footnote{For technical reasons one sometimes needs an
even stronger  restriction, namely {\em superstability}:  a potential 
is {\em superstable} if there are 
two constants $a> 0$ and $b>0$ such that:
$$
\Phi({\bf q}_1, \ldots, {\bf q}_N) \geq - b \, N + \frac{a N^2}{|\Lambda|}
$$
for all ${\bf q}_i \in \Lambda$. A typical example is the Lennard-Jones
potential.} due to the 
accumulation of particles in arbitrarily small regions of space, as a result 
of a too strong short-range attraction (see also subsection
3.2.4). Temperedness assures a sort of ``localizability'' of the interaction
by avoiding a too slow long-range decay. 

Stability and temperedness are satisfied by  the Lennard-Jones potential, 
however, the  important cases of the Coulomb and gravitational potentials 
do not satisfy these requirements.  This situation is partially mitigated by 
superposing a (purportedly more realistic) {\em hard-core} 
potential to them. That is,  a  potential such that 
$\varphi(r) \rightarrow +\infty$ as $r \rightarrow a+$,
( being smooth otherwise), where $a$ is the particle's diameter.
Now, as particles are kept a minimum distance apart, stability is restored,
but not temperedness. Another potential satisfying the requirements,
and which is amenable to calculations, is the so-called  {\em hard-sphere}
potential, describing billiard ball particles (freely moving particles 
interacting only through elastic collisions). Temperedness is automatic as 
this is a finite-range potential. The exception of gravitational and 
electrostatic potentials might signal a different (and more complex) 
thermodynamic behavior for such systems.~\footnote{See, for example the 
odd thermodynamical behavior of stars and, more spectacularly, of 
black-holes.}

\subsubsection{The Canonical Ensemble}

This is the ensemble  describing  a system in contact with a heat reservoir 
at a fixed temperature. Each element of the ensemble is a probability measure 
$\displaystyle \P^{\mbox{\scriptsize can}}_{\Lambda, N, \beta}$, for 
$\beta > 0$, whose density with respect to Lebesgue measure is
$$
\frac{1}{Z_{\Lambda, N, \beta}} \, e^{ -\beta H({\sf q}, {\sf p})},
$$
where the canonical partition function is
$$
Z_{\Lambda, N, \beta} = \int_{\Omega_{\Lambda,N}} \,  
e^{ -\beta H({\sf q}, {\sf p}) }\, \frac{1}{N!} \, 
\displaystyle\Pi_{i=1}^N d^3{\bf q}_i \,  d^3{\bf p}_i.
$$

It can be checked that
$$
T \equiv \frac{2}{3k} 
<\frac{K}{N}>^{\mbox{\scriptsize can}}_{{\Lambda, N, \beta}}
= \frac{2}{3k} <\frac{1}{N} 
\sum_{i=1}^N 
\frac{{\bf p}_i^2}{2m_i}>^{\mbox{\scriptsize can}}_{{\Lambda, N, \beta}}
= \frac{1}{k \, \beta}, 
$$
so that the parameter $\beta$ is essentially the inverse absolute temperature.
Also, the average internal energy $U$ and the average pressure $p$
are given by~\cite{Lof}
$$
U = -\frac{\partial \ln Z_{\Lambda, N, \beta}}{\partial \beta} 
$$
and
$$
p = \frac{1}{\beta} \frac{\partial \ln Z_{\Lambda, N, \beta}}{\partial V}.
$$
Curiously, it turns out that the canonical 
ensemble is orthodic, even without  taking  the thermodynamic limit. One can
then verify  that the thermodynamic free energy $F = U - T \, S$, is
given by $\displaystyle F = F_N(\beta,\Lambda) = - \frac{1}{\beta} \, \ln Z_{\Lambda, N, \beta}$.

The thermodynamic limit  can also be performed for this ensemble, under the
stability and temperedness conditions. So, for example one can prove the 
existence of the specific canonical free-energy in the thermodynamic limit:
$$
f_{can}(\beta,v) = \lim_{\Lambda \uparrow \RR^3, 
\frac{V}{N} \rightarrow v}
\frac{F_N(\beta,\Lambda)}{N},
$$
in terms of which many quantities can be calculated, e.g.,  the
canonical specific internal energy $u_{can} = \frac{\partial 
\beta f_{can}}{\partial
\beta}(\beta,v)$, as well as the canonical pressure $p_{can}$, specific 
entropy $s_{can}$, specific volume and the temperature. 

\subsubsection{The Grand-Canonical Ensemble}

While the two  previous ensembles dealt with systems with a fixed 
total number of particles, the  {\em grand-canonical 
ensemble}  describes a system in  a region $\Lambda$, with fixed 
temperature, but with variable number of 
particles. The phase-space  is now $\Omega_{\Lambda} = 
\cup_{N \geq 0} \Omega_{\Lambda,N}$, where
$\Omega_{\Lambda,N}$ is the set of states with exactly $N$ particles; in 
particular $\Omega_{\Lambda,0}$ consists of only one point: the empty 
(no-particle) or ``vacuum'' state.

The reference measure $\lambda$ is such that for 
any measurable set $A$, we have  
$\lambda(A) = \sum_{N\geq0} \bar{\lambda}_N(A \cap \Omega_{\Lambda})$, where
by convention $\bar{\lambda}_0 (\Omega_{\Lambda,0}) = 1$. Then,
the grand-canonical ensemble is the family of probability measures
$\displaystyle\P^{\mbox{\scriptsize gc}}_{\Lambda, \beta, \mu}$, 
parametrized by $\beta > 0$ and $\mu \in \RR$, whose density with
respect to $\lambda$ is given by
$$
\frac{1}{Z_{\Lambda,\beta, \mu}} \, e^{-\beta(H(\omega) - 
\mu N_{\Lambda}(\omega))},
$$
where the grand-canonical partition function is 
$$
Z_{\Lambda,\beta, \mu} = \int_{\Omega_{\Lambda}}  e^{-\beta(H(\omega) - 
\mu N(\omega))} \, \lambda(d\omega).
$$
In the above, when the system is in a state $\omega$ with exactly $N$
particles, i.e., $N(\omega) = N_{\Lambda} (\omega) = N$, then the Hamiltonian 
is $H(\omega) = H_{\Lambda,N}(\omega)$. Hence, the partition function can
be written as a series,
\begin{eqnarray*}
Z_{\Lambda,\beta, \mu} = \sum_{N=0}^{\infty} \frac{e^{\beta \mu N}}{N!}
\, \int_{(\Lambda \times \RR^3)^N} e^{-\beta H_{N} ({\sf p},{\sf q})}
\,  \Pi_{i=1}^N d^3{\bf q}_i \,  d^3{\bf p}_i \\=  1 +
\sum_{N=1}^{\infty} \frac{z^N}{N!}
\, \int_{\Lambda^N} e^{-\Phi({\sf q})} \,
 \Pi_{i=1}^N d^3{\bf q}_i = \Xi_{\Lambda,\beta,z},
\end{eqnarray*}
where the integration with respect to the  momentum variables is already 
performed~\footnote{Of course one supposes the potential satisfies
$$
\int_{\Lambda^N} e^{-\Phi({\sf q})} \,
 \Pi_{i=1}^N d^3{\bf q}_i < \infty,
$$
for all $N \geq 1$, so that each term of the series is finite.} and  
$\displaystyle  z = e^{\beta \mu} \, (\frac{2 \pi m}{\beta})^{3/2}$ 
is called ``fugacity'' or ``activity'' (which is  approximately
proportional to the density for dilute gases~\cite{Gall, Griffs, Toda}).

Still, the series above could diverge, in which case 
$\displaystyle\P^{\mbox{\scriptsize gc}}_{\Lambda, \beta, \mu}(N_{\Lambda} < \infty) = 0$,  and hence $\displaystyle\P^{\mbox{\scriptsize gc}}_{\Lambda, \beta, \mu}(N_{\Lambda} = +\infty) =1$. In words, the probability that there is 
an infinite number of particles in $\Lambda$ would be one. In order to avoid 
such a collapse of infinitely many particles on any bounded region of space, 
one requires the convergence of the series, which in turn
depends crucially on the potential. 

In fact, stability is 
a sufficient condition,~\footnote{It is also necessary.~\cite{Ruelle}} because
in this case,
\begin{eqnarray*}
\Xi_{\Lambda,\beta, z} =   1 +
\sum_{N=1}^{\infty} \frac{z^N}{N!}
\, \int_{\Lambda^N} e^{-\Phi({\sf q})} \,
 \Pi_{i=1}^N d^3{\bf q}_i  \leq  1 +
\sum_{N=1}^{\infty} \frac{z^N}{N!}
\, |\Lambda|^N \, e^{N \beta B} = e^{z |\Lambda| e^{\beta B}},
\end{eqnarray*}
which is finite for all $z$. Moreover, it follows that the grand-canonical
partition function  is a real  analytic function of $z$ and $\beta$.

The grand-canonical ensemble is orthodic in the thermodynamic limit, with
the grand-canonical pressure given,
for fixed $\beta > 0$ and $z >0$, by
$$
p_{gc}(\beta, z) = \lim_{\Lambda \uparrow \RR^3} p_{\Lambda} (\beta, z) =
\lim_{\Lambda \uparrow \RR^3} \frac{1}{\beta |\Lambda|}
\ln \Xi_{\Lambda,\beta,z}.
$$
and density 
$$
\rho_{gc}(\beta, z) = \lim_{\Lambda \uparrow \RR^3} \rho_{\Lambda} (\beta, z), 
$$
where $$\rho_{\Lambda} (\beta, z) = \frac{<N_{\Lambda}>^{gc}}{|\Lambda|}
= z \beta \frac{\partial p_{\Lambda} (\beta, z)}{\partial z}.$$
At this point, there arises the natural question about the relation of the 
macroscopic variables, calculated at the thermodynamic limit, say, in
the grand-canonical ensemble, to the  corresponding quantities
evaluated using the microcanonical and canonical ensembles. This is linked 
to the important problem of the {\em equivalence of ensembles} (at the
quantities level), about 
which we limit ourselves to the following brief comments. 

If, according to the 
Boltzmann-Gibbs Principle, one could choose any orthodic ensemble to 
describe the equilibrium behavior 
of a given system, and if one agrees to interpret the thermodynamic limit as a 
procedure to extract information about bulk properties of the system 
(disregarding  boundary effects, inevitable when dealing with
any real, hence finite, physical system), then one would expect that the 
choice of ensemble should not be crucial (except, of course, in 
calculational terms). That is, in the sense that they should describe the 
same thermodynamic behavior of the system under study,
the ensembles should be equivalent (in the thermodynamic limit). This is
indeed the case (in the absence of phase transitions), which is proven
by verifying that the ensembles are related to each other through suitable
re-parametrizations of the basic macroscopic variables.~\cite{Ruelle}

As mentioned before, the  three ensembles discussed above are not the 
only orthodic ensembles available. For example, one can create 
new ensembles by imposing fixed external boundary conditions, say,
by imagining  that there are particles at certain fixed positions  
{\em outside} the region $\Lambda$,  with which the 
particles inside can  interact.~\footnote{The
external particle's  momenta are  not important as the interaction potential
is a function of positions only.  Note also that the external particles  
could be assigned according to a given probability distribution, 
or with periodic boundary conditions, etc.}
The interaction potential of the system inside $\Lambda$ has to
be modified accordingly (see also sec. 4.3.1).

Then, working with the corresponding modified Hamiltonian, one can consider
respectively, the microcanonical, canonical and grand-canonical ensembles with
fixed external boundary conditions (thus the previous examples correspond
to the case of free boundary conditions). Under suitable hypothesis on the
distribution of the external particles, these can be shown to be orthodic in 
the thermodynamic limit (under stability and temperedness).~\cite{Gall}

\section{Thermodynamic  
Limit, Infinite-Volume Measures  and Phase Transitions}

There are many reasons for taking the thermodynamic limit. We have already
seen a strong one, namely, to insure orthodicity of the main ensembles 
and, a fortiori,  their equivalence. That is, in order to correctly describe 
the equilibrium thermodynamics of a fluid from microscopic principles, 
one needs to take the thermodynamic limit. 

In any case, one would have expected the need of some kind of 
limiting procedure,~\footnote{As explicitly recognized by 
Hilbert regarding kinetic theory.~\cite{Corry}} when trying to establish 
(in a mathematically sound way) a bridge between two very different 
descriptions of the same system:  that of the  discrete (or
granular) microscopic world of particles and that of the 
continuous (or homogeneous) macroscopic world of thermodynamics. 
A classical example of this discrete-continuum
transition is found in mathematical analysis: in Cantor's contruction of 
the real number system,  the passage from the discrete (an even dense) set of
rational numbers $\QQ$ to the real number {\em continuum} $\RR$ is
accomplished through classes of equivalence of Cauchy sequences; then
any real number is conceived as a limit of 
rationals.~\footnote{{\em En passant}, the non-standard real numbers 
(hyperreals) can in turn be viewed as certain sequences of real numbers. 
For a discussion of continuity, discreteness and its relations with infinity 
and mathematical models, see Ref. 18.}

 One can also view the need of the thermodynamic limit as reflecting the 
{\em change of scales}  involved in the different descriptions, given the
inherently vague  micro-macro distinction in classical statistical mechanics. 
There, in fact,  a system will  qualify  as ``macroscopic''  basically
when it  consists of a ``very  large number'' of tiny (interacting) 
particles; but exactly how many? The usual order of magnitude is given by 
Avogadro's number which, being so huge,  suggests the radical idea of taking 
the limit of infinitely many particles in infinite volume. As the late 
mathematical-physicist R. Dobrushin observed, 
``infinity is a better approximation to the number $6 . 10^{23}$ than the 
number $100$ ($100 \ll 6 . 10^{23} \approx \infty$)''.~\cite{Dobru}  And, 
curiously, it it is sometimes  combinatorially easier to deal directly with 
infinity (as a unit whole) instead of keeping track of each component of a 
finite but huge system.

Of course, real physical systems have a finite number of particles, usually
restricted to a bounded region. Hence, the thermodynamic limit certainly 
is an {\em idealization} (like so many others in the modeling of physical 
systems), justified as a procedure that allows, in the model at
hand, to obtain an exact and precise treatment of  {\em bulk} properties 
of such many-body systems (i.e., properties which would  be not too sensitive
on the finiteness of the system and of boundary effects).
In particular, it opens up the possibility of studying, in a mathematically 
rigorous way, the very difficult and subtle notion of a
{\em phase transition},  which is arguably  the central problem of 
equilibrium statistical mechanics. 

\subsection{What is a phase transition?}

Generally  speaking, a phase-transition is a qualitative change in the
properties of a macroscopic system when it changes from one to another
of it phases. But what are the  ``phases'' of a substance, e.g., a fluid? 
It is hard to find a precise definition in thermodynamics. Intuitively, they 
are the different homogeneous ``forms'' of that same substance, each  
with its characteristic physico-chemical properties and equation of
state. Or else, they are the different  ``states of aggregation'' of 
matter,~\cite{Somme} an unmistakably  microscopic viewpoint. 

For a fluid,  we have the familiar 
solid,~\footnote{With many different possible crystalline phases.} gas and  
liquid  phases, which are geometrically  described by the set of states
comprising certain {\em sectors} in the $(p,v)$ (or $(p,T)$) state-space or 
phase diagram. These sectors seem to be separated by well-defined 
{\em coexistence curves} where two different phases can coexist at the same 
value of the thermodynamic parameters. Besides, at  such curves the equation 
of state seems to break down due to the appearance of {\em singularities} or,
more specifically, {\em nonanalityticities}, in some 
thermodynamic quantities, like pressure. 

This is a picture corroborated by countless experiments 
(and numerical simulations) and which one would like to 
explain from statistical mechanics. However, this  turned out to be 
an extremely difficult problem and, although there is a very detailed
understanding of it for some lattice systems, is still essentially open 
for continuous models.

 Now, even to start such an ambitious goal, one would surely need
a precise notion of phase transition in the context of statistical mechanics. 
And the fact is that there are,  at present, different  notions of phase 
transitions around, usually suggested  by some fundamental negative
results, that is, concerning the {\em absence} of phase transition (see below).

If one examines  the phase diagram of a fluid system, 
the situation at a point on the coexistence curve seems to indicate that
the thermodynamic parameters (or state variables)
do not  uniquely specify the  equilibrium 
``macrostate'' of the system. It could be, for example, 
liquid or solid at the liquid-solid coexistence curve, with different 
proportions of each phase. Also, the crossing of such curves usually 
manifests itself through some ``abrupt'' 
(for example, discontinuous) change in some thermodynamic quantities. These 
observations are the basis of two popular notions of phase transition,
that we briefly describe next.~\cite{Kote}

\subsection{ Phase transitions as singularities of thermodynamic
potentials}

The idea is  as follows.  Many important 
thermodynamical quantities are obtained as derivatives of a thermodynamic 
potential with respect to the basic parameters of the chosen ensemble.  
Hence, the presence of a discontinuity on some such quantity signals that 
the potential is non-differentiable at some point, i.e., it 
is {\em singular}: such a point (in the parameter space) will be called a 
phase-transition point. 

 In principle, this would provide  a method to pin-point the  
values of the basic parameters at which a phase-transition occurs. One 
then loosely define  a phase-transition as a singularity of the 
thermodynamic potential (due, for example to the discontinuity or non-existence
of some of its derivatives).

 However, in finite volume the thermodynamic potentials are smooth functions 
of the basic parameters (being given  as expectation values of the partition 
function). We have seen, for example, that the finite-volume 
grand-canonical partition function is a real analytic function of the basic 
parameters. Therefore, one needs to take the thermodynamic limit if one 
hopes to observe the appearance of a  singularity. This provides yet another 
justification for taking the thermodynamic limit: it is needed 
in order to be able to have a sharp (mathematically precise) manifestation 
of a phase-transition. 

In this way one would hope to study the structure of
parameter-space (or phase diagram) by, say, separating  the regions where 
there is or not a phase-transition.  This approach has been more succesfull 
in providing proofs of  {\em absence} of phase transitions. So, for example,  
there are classical results~\cite{Gall, Griffs, Ruelle} showing that in the
thermodynamic limit the grand-canonical pressure $p_{gc}(\beta,z)$ is an 
analytic function of $(\beta, z)$ for sufficiently small values of 
inverse temperature $\beta >0$ or of fugacity $z >0$ (and for these 
so-called {\em regular} values the equivalence of ensembles holds). In other 
words, for sufficiently high temperatures and/or sufficiently low densities, 
there is no phase-transition. 

The main defect  of this approach is that it provides
no clear physical mechanism to explain the appearance of the singularities.
However, as at those values of the parameters the system would presumably
be in the gas phase, there is at least a hint that particles would be
so far apart that they could not interact strongly enough to 
begin forming  ``aggregates'' (or ``clusters'') which would eventually lead 
to the condensation process.

\subsection{Phase transitions as non-uniqueness of infinite volume
measures}

This  alternative approach to the description of phase transition 
is inspired by the above mentioned non-uniqueness of 
the ``macrostate'' at a coexistence curve. The precise formulation,
however, is much more abstract: first of all, it proposes to work directly
in an infinite-volume setting, leading  to  the notion of the 
thermodynamic limit {\em at the level of (probability) measures}.
 
At first, this is just an extension of the thermodynamic limit procedure
(discussed in the last section  for some specific quantities) 
to the whole set of {\em local} state-variables. Recall  that a 
state-variable is a  measurable function (say, bounded or integrable) 
$F: \Omega_{\Lambda} \rightarrow \RR$ on phase-space.

Let $\Delta \subset \Lambda$ be an  open bounded set. Then 
$F$ is said to be {\em localized}  in $\Delta$  if it does not depend on 
position and momentum coordinates of particles lying outside of 
$\Delta$ (examples are kinetic energy, potential energy, etc).

Consider, in the grand-canonical ensemble (with fixed $\beta$ and 
$\mu$), for each local state-variable $F$,  the limit, 
$$
<F>^{\mbox{\scriptsize gc}}_{\beta, \mu} \equiv 
\lim_{\Lambda \uparrow \RR^3} <F>^{\mbox{\scriptsize gc}}_{\Lambda, \beta, \mu} = 
\lim_{\Lambda \uparrow \RR^3} \int_{\Omega_{\Lambda}} F(\omega) \, 
\P^{\mbox{\scriptsize gc}}_{\Lambda, \beta, \mu}(d\omega),
$$
for a suitable increasing sequence of space-filling volumes. Under certain 
restrictions on the potential (i.e., superstability) it is possible
to use standard compactness arguments to prove that such limits exist, 
at least along certain subsequences.~\cite{Lan, Pulv, Pet}
Moreover, if they  exist,  one can show (using a version of
the Riesz-Markov theorem) that 
the $<F>^{\mbox{\scriptsize gc}}_{\beta, \mu}$, for all local $F$,
determine a unique probability measure 
$\displaystyle\P^{\mbox{\scriptsize gc}}_{\beta, \mu}$ on
a certain {\em infinite-volume phase-space}  $\Omega$, with
$$
<F>^{\mbox{\scriptsize gc}}_{\beta, \mu} = 
\int_{\Omega} F(\omega) \, 
\P^{\mbox{\scriptsize gc}}_{\beta, \mu}(d\omega),  
$$
so that they are  expectations with respect to that measure.

Such probability measure is called an 
{\em infinite-volume limit (or cluster) measure}. 
There is an associated notion of (weak) convergence on the space of probability 
measures  on $(\Omega, \mathcal M)$, such that all the above can 
be summarized  by saying that cluster measures are (weak) limits 
(as $\Lambda \uparrow \RR^3$)  of the corresponding finite-volume
grand-canonical  measures, thus: 
$\displaystyle \P^{\mbox{\scriptsize gc}}_{\Lambda, \beta, \mu} 
\Rightarrow \P^{\mbox{\scriptsize gc}}_{\beta, \mu}$. 

  Of course, there are many technical 
details involved here. To begin with, one needs to describe what is the 
infinite-volume  phase-space $\Omega$. It will consist of all symmetrized 
(i.e., permutation-invariant) and  {\em locally finite} sequences of 
particle's position and momenta, the last requirement meaning  that  
only a {\em finite} number of particles are allowed in any open  bounded 
subset  $\Lambda \in \RR^3$.~\footnote{That is, for $\omega \in \Omega$,
$\omega = \{({\bf q}_i, {\bf p}_i)\}_{\{i\geq 1\}}$, then for
any bounded open set $\Lambda \in \RR^3$, we have 
$\mbox{card}\{\omega_{\Lambda}\} < \infty$, where 
$\omega_{\Lambda} = \omega \cap (\Lambda \times \RR^3)$, and 
$\mbox{card}\{A\}$ 
means the cardinality of the set $A$. The space $\Omega$ is endowed with the
topology of local convergence: a sequence 
$\omega_n = \{({\bf q}^n_i, {\bf p}^n_i)\}_{\{i\geq 1\}}$ converges to
$\omega = \{({\bf q}_i, {\bf p}_i)\}_{\{i\geq 1\}}$ if  
$\displaystyle\lim_{n \rightarrow \infty} {\bf q}^n_i = {\bf q}_i$ and  
$\displaystyle\lim_{n \rightarrow \infty} {\bf p}^n_i = {\bf p}_i$, for some
enumeration of position and momenta. More precisely, such that for 
all bounded open $\Lambda$ such that
$\mbox{card}\{\omega \cap \partial(\Lambda \times \RR^3)\}$, there exists
an $n_0$ such that for all $n \geq n_0$ it holds that 
$\mbox{card}\{\omega_n \cap (\Lambda \times \RR^3)\} = 
\mbox{card}\{\omega \cap (\Lambda \times \RR^3)\}$. 

Consider the natural projection $\pi_{\Lambda} : \Omega \rightarrow 
\Omega_{\Lambda}$, with $\pi_{\Lambda}(\omega) = \omega_{\Lambda}$. Then,
a state-function $F$ is localized in $\Lambda$ if $F(\omega) = 
F(\omega')$ for all $\omega$, $\omega'$ such that 
$\pi_{\Lambda}(\omega) = \pi_{\Lambda}(\omega')$.}

The above discussion was based on choosing the (finite volume free boundary)
grand-canonical ensemble, and one could ask what happens if
one begins with a different ensemble (possibly including those with
boundary condition). This brings up again the question of the equivalence of
ensembles, now  {\em at the level of measures} which was  recently
dealt with rigorously.~\cite{Giorg}

\subsubsection{The DLR-equation}

At this point, one should mention yet another,
more general and very  elegant  (and much less known) viewpoint, not directly
involving limits: the so-called DLR equation. 
It is motivated by the 
following semi-rigorous reasoning.~\cite{Pulv}
 
Let $\nu_{\Lambda}$ denote the finite-volume grand-canonical measure 
(where, for simplicity, we do not write the parameters $\beta$ and
$\mu$), that is: 
$$
\nu_{\Lambda}(d\omega) = \frac{1}{Z_{\Lambda}} \, 
e^{-\beta(H(\omega) - \mu N(\omega))} \, \lambda(d\omega).
$$

Then, for any $\Delta \subset \Lambda$, we can identify  the space
$\Omega_{\Lambda}$ with the cartesian  product 
$\Omega_{\Delta} \times \Omega_{\Delta^c}$ (where 
$\Delta^c = \Lambda - \Delta$), each state being denoted
by $\omega = \omega_{\Lambda} = \{\omega_{\Delta}, \omega_{\Delta^c}\}$.
Then the  reference measure $\lambda$ can be  identified with the 
product measure $\lambda_{\Delta} \otimes \lambda_{\Delta^c}$.

The  Hamiltonian in $\Omega_{\Lambda}$ is then written as
$$
H(\omega_{\Lambda}) = H(\omega_{\Delta}) + H(\omega_{\Delta^c}) +
W(\omega_{\Delta}|\omega_{\Delta^c}),
$$
where
$$
W(\omega_{\Delta}|\omega_{\Delta^c}) = \sum_{({\mathbf q}_i, {\mathbf p}_i) 
\in \omega_{\Delta}} \;  \sum_{({\mathbf q}_j, {\mathbf p}_j) \in 
\omega_{\Delta^c}}  \varphi(|{\mathbf q}_i - {\mathbf q}_j|),
$$
is the potential energy of interaction of particles inside $\Delta$  with
particles outside of it. 

We then have,
\begin{multline*}
\nu_{\Lambda}(d\omega) =  \nu_{\Lambda}(d\omega_{\Delta}, d\omega_{\Delta^c})
\,
= \\ \frac{1}{Z_{\Lambda}} \, e^{-\beta(H(\omega_{\Delta^c}) -
\mu N(\omega_{\Delta^c}))} \,  e^{-\beta(H(\omega_{\Delta}) + 
W(\omega_{\Delta}|\omega_{\Delta^c}) - \mu N(\omega_{\Delta}))} \, 
\lambda_{\Delta}(d\omega_{\Delta}) \otimes 
\lambda_{\Delta^c}(d\omega_{\Delta^c}).
\end{multline*}

By  Fubini's theorem,  for any bounded measurable 
state-function $F$ on 
$\Omega_{\Lambda}$, we have
\begin{multline*}
\int_{\Omega_{\Lambda}} \nu_{\Lambda}(d\omega_{\Lambda}) \, 
F(\omega_{\Lambda})  
 = \\
\int_{\Omega_{\Delta^c}}  \, \lambda(d\omega_{\Delta^c})  
\, e^{-\beta(H(\omega_{\Delta^c}) - \mu N(\omega_{\Delta^c}))} 
\frac{Z_{\Delta}(\omega_{\Delta^c})}{Z_{\Lambda}} \,  
\int_{\Omega_{\Delta}} \, g(d\omega_{\Delta}|\omega_{\Delta^c})
F(\omega_{\Delta}, \omega_{\Delta^c}),
\end{multline*}
where $g(\cdot|\omega_{\Delta^c})$ (sometimes called a 
{\em Gibbs specification}) is just the finite-volume grand-canonical
probability measure on $(\Omega_{\Delta}, {\mathcal M}_{\Delta})$, with 
boundary conditions $\omega_{\Delta^c}$.  That is,
$$
g(d\omega_{\Delta}|\omega_{\Delta^c}) = 
\frac{1}{Z_{\Delta}(\omega_{\Delta^c})} \, e^{-\beta(H(\omega_{\Delta}) + 
W(\omega_{\Delta}|\omega_{\Delta^c}) - \mu N(\omega_{\Delta}))}\, 
\lambda_{\Delta}(d\omega_{\Delta}),
$$
with corresponding partition function $Z_{\Delta}(\omega_{\Delta^c})$.

 In sum, we have:
\begin{eqnarray*}
\int_{\Omega_{\Lambda}}  \nu_{\Lambda}(d\omega_{\Lambda}) \, 
F(\omega_{\Lambda})  = 
\int_{\Omega_{\Delta^c}}  \, 
\nu_{\Lambda}( \Omega_{\Delta}, d\omega_{\Delta^c})  
\int_{\Omega_{\Delta}} \, g(d\omega_{\Delta}|\omega_{\Delta^c})
F(\omega_{\Delta}, \omega_{\Delta^c}),
\end{eqnarray*}
where we used that 
$\displaystyle\nu_{\Lambda}(\Omega_{\Delta}, d\omega_{\Delta^c}) =
\int_{\Omega_{\Delta}} \nu_{\Lambda}(d\omega_{\Delta}, d\omega_{\Delta^c})$.

Having in mind the infinite-volume limit, $\Lambda \uparrow \RR^3$, 
this suggests the following definition: a probability
measure $\P$ in  $(\Omega, \mathcal M)$
is called an {\em infinite-volume Gibbs measure}  (or distribution)
with interaction potential $\varphi$, inverse temperature  $\beta$ and 
chemical  potential $\mu$ if, for every bounded set $\Delta \in \RR^3$ and
all localized functions $F$, it satisfies  the so-called 
{\em DLR-equation} (after Dobrushin, Lanford and 
Ruelle):
$$
\int_{\Omega} \P(d\omega) F(\omega) = \int_{\Omega} \P(d\omega) \,
\int_{\Omega_{\Delta}} g(d\omega_{\Delta}|\omega_{\Delta^c}) \, F(\omega).
$$
Now, 
$\Delta^c = \RR^3 - \Delta$ and we identify $\Omega = \Omega_{\Delta}
\times \Omega_{\RR^3 - \Delta}$.~\footnote{An equivalent formulation is as 
follows: a probability measure $\P$ on $(\Omega, \mathcal M)$ is an 
infinite-volume Gibbs measure with interaction potential $\Phi$ and 
parameters $(\beta, \mu)$ if, for any bounded $\Lambda \in \RR^3$:
\begin{itemize} 
\item[(i)] for $\P$-almost  every $\omega \in \Omega$, there exists
the grand-canonical distribution with interaction potential $\varphi$ and
parameters  $(\beta, \mu)$ ,  in the (finite) volume $\Delta$ and with
boundary conditions $\omega_{\Delta^c}$ (in other words the
partition function $Z_{\Delta} (\omega_{\Delta^c}) < \infty$, $\P$-almost
everywhere);
\item[(ii)] let $\P(\cdot|{\mathcal M}_{\Delta^c})$ be (a version of)
the conditional probability distribution of of $\P$ with respect to the 
$\sigma$-algebra ${\mathcal M}_{\Delta^c}$; then, for $\P$-almost  every 
$\omega \in \Omega$, its restriction to ${\mathcal M}_{\Delta}$ is
absolutely continuous with respect to the reference measure $\lambda$,
with density (given by the Radon-Nikodym derivative):
$$
p_{\Delta}(\omega_{\Delta}|\omega_{\Delta^c}) = 
\frac{dg(\cdot|\omega_{\Delta^c})}{d\lambda}(\omega_{\Delta}) = 
\frac{1}{Z_{\Delta}(\omega_{\Delta^c})} \, e^{-\beta(H(\omega_{\Delta}) + 
W(\omega_{\Delta}|\omega_{\Delta^c}) - \mu N(\omega_{\Delta}))}.
$$
\end{itemize}
The (i) and (ii) are called the {\em DLR-conditions}.}

 Although a bit technically complicated, the idea is quite straightforward: 
an infinite-volume Gibbs measure is such that, when conditioned on events 
{\em outside}  any given bounded region $\Delta$ and then restricted to events 
on $\Delta$, we get exactly  a finite-volume grand-canonical distribution, 
with the corresponding boundary condition.

Under certain technical assumptions on the potential, it is known
that every infinite-volume limit  measure (in the sense  discussed
in the  previous section)  is a solution of the DLR-equation  and, 
conversely, every infinite-volume Gibbs measure is the infinite-volume 
limit of a finite-volume grand-canonical  measure with some (random) 
boundary conditions (for the delicate and difficult proofs of these results, 
see Ref. 46).  Moreover, there always exists  a solution of  the 
DLR-equations. 

It is quite possible that, for a given pair $(\beta, \mu)$, there exists
more than one solution to the DLR-equation. However, it is 
proven~\cite{Dob,Pet} that at sufficiently high temperature or low
density  there exists a {\em  unique}  solution of DLR-equations, which
is translation-invariant (which is important because such measures would be 
interpreted as the ``pure'' phases of the macroscopic system). Moreover,
this unique solution has exponential decay of correlations, which would mean
that particles do not tend to form ``clusters'', supposedly the mechanism
working in gas condensation.~\footnote{Incidentally, the
property of exponential decay of correlations is yet another
characterization of absence of phase transition found in the literature.}
In conjunction with the analyticity properties
of the thermodynamic potential, these results characterize the
{\em absence} of phase-transition for that range of the parameters
$(\beta, \mu)$. 

Correspondingly, at those values of the parameters for  which there exist
more than one solution of the DLR-equation, a phase transition is said
to occur. That is, the  {\em non-uniqueness of the infinite-volume Gibbs 
measure} is taken to signal the occurrence of a  phase-transition. For 
example, if $(\beta, \mu)$ belongs to the liquid-vapor coexistence line, one  
would expect the existence of only two {\em extremal} translation-invariant 
Gibbs measures, 
$\P_l$  and $\P_g$.~\footnote{For  a variational 
characterization of such measures, see Ref. 34.}
This is interpreted by saying that  these measures describe the ``pure'' 
liquid  and  gas phases (respectively),  so that  any other 
translation-invariant Gibbs measure $\P^{(b)}$,  with  boundary conditions 
denoted by $b$, is a convex 
combination of them, i.e., 
$$
\P^{(b)} = \alpha \, \P_l + (1-\alpha) \, \P_g,
$$
where  $\alpha \in [0,1]$ would depend on the boundary conditions $b$. 

Each such $\P^{(b)}$ is a interpreted as a ``mixture'' of phases at 
coexistence, with clusters (maybe drops) of liquid amidst vapor and 
$\alpha$ being  the ``proportion'' of liquid phase,
$1-\alpha$ that of the gas phase.   As this proportion depends on the
boundary conditions $b$, it appears  that a phase transition can also
be viewed as a kind of {\em instability}  of the system, which becomes 
sensitive to the (infinite-volume) boundary condition chosen; 
in other words, it becomes highly  
correlated.~\footnote{There could also exist non-translation invariance
Gibbs measures, which would correspond to phase coexistence favoring the
formation of a separating interface.}

While this  kind of scenario is basically proven in the case, for example,
of  the Ising model,~\cite{Lebo}  there is no corresponding results
for continuous systems.  That is, the problem of proving the {\em existence}
of phase transitions for fluids, by showing the non-uniqueness of the 
infinite-volume Gibbs measures  at suitable  parameter values, is an 
essentially open problem. Only very recently there was a 
breakthrough,~\cite{Pre} with  a proof of existence of the liquid-vapor 
phase transition for a continuous particle  model interacting through a 
finite-range Kac-type potential.~\footnote{There is yet no corresponding 
proof for the case of Lennard-Jones potential,~\cite{Lebo} not to mention the 
question of proving the existence of a crystalline (solid) 
phase.~\cite{Ruelle3}}

 We end this discussion by realizing that at present there is no
consensus on what is (or should be) the appropriate 
definition of a phase transition,~\footnote{We should also mention the
critical exponents viewpoint, a more phenomenological approach
with a huge literature, and which tries to
describe and classify the singular behavior of 
quantities close to a phase transition, with the associated notions 
of universality, scaling,
renormalization, etc.~\cite{Domb} There is also a topological
viewpoint of phase transitions, see ref. 8} and this is probably due 
to the fact that one does not quite understand the physical phenomenon itself.
Note also that there is not a clear and complete understanding of the 
relationships among the different notions currently in use by physicists and
mathematical-physicists.

\section{Conclusions}

In this paper we tried to  examine some basic notions behind the
structure of classical equilibrium statistical mechanics.
We argued that statistical mechanics was born as a
level-connecting discipline, in the specific context of the attempts to
provide a mechanical-atomistic explanation of thermodynamics.

At least for the equilibrium case, the micro-macro link is effected through 
the  Boltzmann-Gibbs prescription,  with the help of additional hypothesis 
such as the the thermodynamic limit. At the formal (mathematical) level 
this is accomplished  through the crucial level-linking concept of 
the ensembles, that is, families of invariant probability measures on 
the microscopic phase-space, indexed by the macroscopic parameters. 
Probabilistic methods and notions (old and new) are essentially present, 
but they are not necessarily associated to any random mechanisms.

Incidentally, the need of  such ``extra'' hypothesis as the thermodynamic 
limit, shows  that the ``reduction''  of thermodynamics to statistical 
mechanics is not a simple matter. It requires
the development of sophisticated mathematical-physical concepts and 
techniques, specially of a probabilistic sort, such as the notion of 
infinite volume Gibbs measures and the DLR-equation.  Moreover, as the 
delicate and complicated issue of phase transitions shows, the statistical 
mechanical program is far from being completed, in spite of some enormous 
advances.

On the other hand, the  very success of the ensemble method of equilibrium 
statistical mechanics have inspired its application not only to the study of 
many-body classical and quantum mechanics, but to many other fields
dealing with systems with many interacting ``microscopic'' 
components from of which one hopes to deduce some corresponding ``macroscopic''
behavior through an averaging procedure.

The reasons for this ``portability'' of statistical mechanical
methods, given the somewhat restricted context to which it was originally
linked, are not quite clear. A crucial ingredient  surely is
the central role of probability theory in its framework, 
with its unifying language, methods and results. Another would be
the  relative simplicity of the recipe to be followed in such applications, 
which boils down to: in studying a system with a very large number
of similar interacting components ``in equilibrium'', apply the 
Boltzmann-Gibbs prescription, with a suitable Hamiltonian, a suitable notion 
of temperature, etc, and try  to derive the consequences.  This does not in
any way mean that it is an easy task to derive useful, not to mention, 
meaningful results from this procedure.

But what  justifies the use of the Boltzmann-Gibbs prescription, besides its
practical successes.  In many of the applications outside the original 
thermodynamic systems,  different concepts of ``entropy''  are usually 
introduced, loosely interpreted as measuring ``disorder'', and the 
Boltzmann-Gibbs prescription is ``justified'' by a variational principle 
which requires that the entropy should be maximized. Though such justifications
might be satisfactory as far as some of these application go, and  though
there are variational principles also in the standard 
statistical mechanics, these will not provide an
explanation for the Boltzmann-Gibbs principle from first principles.

  It is an old dream of one of the founders of the field, Ludwig
Boltzmann, that the ultimate  justification of equilibrium statistical
mechanics would lie at a  deeper level, namely,
at the basic non-equilibrium dynamics of the system. That is, one should 
somehow derive equilibrium statistical mechanics from a 
(still non-existent!) theory of non-equilibrium statistical mechanics. In 
spite of some important advances in this area, it still  remains the basic 
foundational open problem of statistical mechanics and of physical science.

\bigskip

\bigskip

\bigskip
\noindent{\bf ACKNOWLEDGMENTS} 

\bigskip
This work was partially supported by
FAPERJ, Projeto Cientista do Nosso Estado, E-26/151.905/2000.
\bigskip


\vspace{-2mm}
\noindent\hrulefill

\bigskip


\end{document}